\documentclass[twocolumn,trackchanges]{aastex61}

\newcommand{\icnu}{IceCube-170922A}
\newcommand{\txs}{TXS~0506+056}
\newcommand{\enu}{\mbox{$\varepsilon_\nu$}}
\newcommand{\egam}{\mbox{$\varepsilon_\gamma$}}
\newcommand{\ecr}{\mbox{$\varepsilon_{\rm cr}$}}
\newcommand{\epro}{\mbox{$\varepsilon_p$}}

\newcommand{\pcharge}{\mbox{\texttt{charge}}}
\newcommand{\penergy}{\mbox{\texttt{energy}}}

\newcommand{\psig}{\mbox{\texttt{signalness}}}

\newcommand{\amonehe}{\mbox{\texttt{AMON\_ICECUBE\_EHE}}}
\newcommand{\eFenu}{\mbox{$\varepsilon_\nu F_{\varepsilon_\nu}$}}
\newcommand{\eFegam}{\mbox{$\varepsilon_\gamma F_{\varepsilon_\gamma}$}}
\newcommand{\eFenumax}{\mbox{$\varepsilon_\nu F^{(\max)}_{\varepsilon_\nu}$}}
\newcommand{\emaillink}[1]{\href{mailto:#1}{\texttt{#1}}}

\newcommand{\dg}{$^\circ$}

\newcommand{\ice}{\mbox{IceCube}}
\newcommand{\swift}{\textit{Swift}}
\newcommand{\nustar}{\textit{NuSTAR}}
\newcommand{\xshooter}{\mbox{X-shooter}}
\newcommand{\fermi}{\textit{Fermi}}
\newcommand{\etal}{\mbox{et al.}}
\newcommand{\xray}{\mbox{X-ray}}
\newcommand{\xrays}{\mbox{X-rays}}
\newcommand{\gray}{\mbox{$\gamma$-ray}}
\newcommand{\grays}{\mbox{$\gamma$-rays}}

\newcommand{\percmsq}{\mbox{cm$^{-2}$}}

\newcommand{\ergcms}{\mbox{erg cm$^{-2}$ s$^{-1}$}}
\newcommand{\ergsec}{\mbox{erg s$^{-1}$}}
\newcommand{\ctsec}{\mbox{ct s$^{-1}$}}
\newcommand{\ctksec}{\mbox{ct ks$^{-1}$}}

\newcommand{\rah}{\mbox{$^{\rm h}$}}
\newcommand{\ram}{\mbox{$^{\rm m}$}}
\newcommand{\ras}{\mbox{$^{\rm s}$}}
\def\fs{\hbox{$.\!\!^{\rm s}$}}
\newcommand{\rasxg}[3]{\mbox{#1\rah\,#2\ram\,#3\ras}}
\newcommand{\rasxgd}[4]{\mbox{#1\rah\,#2\ram\,#3\fs #4}}
\def\arcdeg{\hbox{$^\circ$}}
\def\arcmin{\hbox{$^\prime$}}
\def\arcsec{\hbox{$^{\prime\prime}$}}
\def\farcs{\hbox{$.\!\!^{\prime\prime}$}}
\newcommand{\dcsxg}[3]{\mbox{#1\arcdeg\,#2\arcmin\,#3\arcsec}}
\newcommand{\dcsxgd}[4]{\mbox{#1\arcdeg\,#2\arcmin\,#3\farcs #4}}
\newcommand{\eqb}{\begin{equation}}
\newcommand{\eqe}{\end{equation}}

\def\gtrsim{\ {\raise-.5ex\hbox{$\buildrel>\over\sim$}}\ }
\def\lesssim{\ {\raise-.5ex\hbox{$\buildrel<\over\sim$}}\ }
\def\simlt{\mathrel{\hbox{\rlap{\hbox{\lower4pt\hbox{$\sim$}}}\hbox{$< $}}}}
\def\simgt{\mathrel{\hbox{\rlap{\hbox{\lower4pt\hbox{$\sim$}}}\hbox{$> $}}}}

\def\aap{\ {A\&A}\ }

\def\aj{\ {AJ}\ }

\def\apj{\ {ApJ}\ }
\def\apjl{\ {ApJL}\ }
\def\apjs{\ {ApJS}\ }

\def\mnras{\ {MNRAS}\ }

\def\pasp{\ {PASP}\ }

\def\ssr{\ {Space Sci. Rev.}\ }

\makeatletter

\def\frontmatter@affiliationfont{\normalfont\scriptsize\it
\ifmodern\baselineskip=6pt\fi
\iflongauthor\else
\rightskip-12pt plus 1fil
\leftskip6pt \parindent-4pt
\fi
}%

\def\@affil@script#1#2#3#4{%
\iffirstaffil\vskip6pt 
\global\firstaffilfalse\fi
 \@ifnum{#1=\z@}{}{%
  \par
  \begingroup
   \frontmatter@affiliationfont
   \@ifnum{\c@affil<\affil@cutoff}{}{%
\def\one{#1}
\ifnum\one<\largestAffilNum
   \def\@thefnmark{#1}\@makefnmark\fi
\ifnum\one=\largestAffilNum
   \def\@thefnmark{#1}\@makefnmark\fi
   }%
\ifnum\one<\largestAffilNum
   \ignorespaces#3%
\fi
\ifnum\one=\largestAffilNum
   \ignorespaces#3%
\fi
   \@if@empty{#4}{}{\frontmatter@footnote{#4}}%
   \par
  \endgroup
 }%
}%

\makeatother

\renewcommand{\edit}[2]{#2}

\begin{document}


\title{A Multimessenger Picture of the Flaring Blazar TXS 0506+056:\\
Implications for High-Energy Neutrino Emission and Cosmic Ray Acceleration}


\correspondingauthor{Azadeh Keivani, Kohta Murase,
  Maria Petropoulou, Derek Fox}
\email{keivani@psu.edu, murase@psu.edu,
  m.petropoulou@astro.princeton.edu, dfox@psu.edu}


\author{A. Keivani}
\affil{Department of Physics,
Pennsylvania State University,
University Park, PA 16802, USA}
\affil{Center for Particle \& Gravitational Astrophysics,
Institute for Gravitation and the Cosmos,
Pennsylvania State University,
University~Park, PA 16802, USA}

\author{K. Murase}
\affil{Department of Physics, Pennsylvania State University,
  University Park, PA 16802, USA}
\affil{Center for Particle \& Gravitational Astrophysics,
  Institute for Gravitation and the Cosmos,
  Pennsylvania State University,
  University~Park, PA 16802, USA}
\affil{Department of Astronomy \& Astrophysics,
  Pennsylvania State University,
  University Park, PA 16802, USA}
\affil{Center for Gravitational Physics, Yukawa Institute for
  Theoretical Physics, Kyoto, Kyoto 606-8502 Japan}

\author{M. Petropoulou}
\affil{Department of Astrophysical Sciences,
Princeton University, Princeton, NJ 08544, USA}

\author{D.~B. Fox}
\affil{Center for Particle \& Gravitational Astrophysics,
  Institute for Gravitation and the Cosmos,
  Pennsylvania State University,
  University~Park, PA 16802, USA}
\affil{Department of Astronomy \& Astrophysics,
  Pennsylvania State University,
  University Park, PA 16802, USA}
\affil{Center for Theoretical \& Observational Cosmology,
  Institute for Gravitation and the Cosmos, 
  Pennsylvania State University, University Park, PA 16802, USA}


\author{S.~B. Cenko}
\affil{Astrophysics Science Division, NASA Goddard Space Flight Center, Mail Code 661, Greenbelt, MD 20771, USA} 
\affil{Joint Space-Science Institute, University of Maryland, College Park, MD 20742, USA}

\author{S. Chaty}
\affil{Laboratoire AIM (UMR 7158 CEA/DRF - CNRS - Universit\'e Paris Diderot), Irfu / D\'epartement d'Astrophysique, CEA-Saclay, FR-91191 Gif-sur-Yvette Cedex, France} 

\author{A. Coleiro}
\affil{APC, Univ Paris Diderot, CNRS/IN2P3, CEA/Irfu, Obs de Paris, Sorbonne Paris Cit\'e, France} 
\affil{IFIC - Instituto de F\'isica Corpuscular (CSIC - Universitat de Val\`encia), Calle Catedr\'atico Jos\'e Beltr\'an, 2 E-46980 Paterna, Valencia, Spain}


\author{J.~J. DeLaunay}
\affil{Department of Physics,
  Pennsylvania State University,
  University Park, PA 16802, USA}
\affil{Center for Particle \& Gravitational Astrophysics,
  Institute for Gravitation and the Cosmos,
  Pennsylvania State University,
  University~Park, PA 16802, USA}

\author{S. Dimitrakoudis}
\affil{Department of Physics, 
  University of Alberta, 
  Edmonton, Alberta T6G 2E1, Canada}

\author{P.~A. Evans}
\affiliation{Department of Physics \& Astronomy,
  University of Leicester,
  Leicester, LEI 7RH, UK}

\author{J.~A. Kennea}
\affil{Department of Astronomy \& Astrophysics,
  Pennsylvania State University,
  University Park, PA 16802, USA}

\author{F.~E. Marshall}
\affil{NASA Goddard Space Flight Center, Mail Code 660.1, Greenbelt, MD 20771, USA} 

\author{A. Mastichiadis}
\affil{Department of Physics, 
  National and Kapodistrian University of Athens, 
  Panepistimiopolis, GR 15783 Zografos, Greece}
 
\author{J.~P. Osborne}
\affil{Department of Physics \& Astronomy,
  University of Leicester,
  Leicester, LEI 7RH, UK}

\author{M. Santander}
\affil{Department of Physics and Astronomy, 
  University of Alabama, 
  Tuscaloosa, AL 35487, USA}

\author{A. Tohuvavohu}
\affil{Department of Astronomy \& Astrophysics,
  Pennsylvania State University,
  University Park, PA 16802, USA}
  
\author{C.~F. Turley}
\affil{Department of Physics,
  Pennsylvania State University,
  University Park, PA 16802, USA}
\affil{Center for Particle \& Gravitational Astrophysics,
  Institute for Gravitation and the Cosmos,
  Pennsylvania State University,
  University~Park, PA 16802, USA}


\begin{abstract}
Detection of the \icnu\ neutrino coincident with the flaring blazar \txs, the first and only $\sim$3$\sigma$ high-energy neutrino source association to date, offers a potential breakthrough in our understanding of high-energy cosmic particles and blazar physics. 
We present a comprehensive analysis of \txs\ during its flaring state, using newly collected \swift, \nustar, and \xshooter\ data with \fermi\ observations and numerical models to constrain the blazar's particle acceleration processes and multimessenger (electromagnetic and high-energy neutrino) emissions. 
Accounting properly for electromagnetic cascades in the emission region, we find a physically-consistent picture only within a hybrid leptonic scenario, with \grays\ produced by external inverse-Compton processes and high-energy neutrinos via a radiatively-subdominant hadronic component. 
\edit1{We derive robust constraints on the blazar's neutrino and cosmic-ray emissions and demonstrate that, because of cascade effects, the 0.1--100\,keV emissions of \txs\ serve as a better probe of its hadronic acceleration and high-energy neutrino production processes than its GeV--TeV emissions.}
If the \ice\ neutrino association holds, physical conditions in the \txs\ jet must be close to optimal for high-energy neutrino production, \edit1{and are not favorable for} ultra-high-energy cosmic-ray acceleration. 
Alternatively, the challenges we identify \edit1{in} generating a significant rate of \ice\ neutrino detections from \txs\ \edit1{may} disfavor single-zone models, in which \mbox{$\gamma$-rays} and high-energy neutrinos are produced in a single emission region. 
\edit1{In concert with continued operations of the high-energy neutrino observatories, we advocate regular \xray\ monitoring of \txs\ and other blazars in order to test single-zone blazar emission models, clarify the nature and extent of their hadronic acceleration processes, and carry out the most sensitive possible search for additional multimessenger sources.}
%
%
\end{abstract}





\section{Introduction}
\label{sec:intro}

High-energy (HE; $\enu\simgt {\rm 1\,TeV}$) neutrinos, as cosmic messenger particles, have the potential to reveal the sources of HE cosmic rays and illuminate their underlying particle acceleration processes. 
Detection of a nearly-isotropic flux of HE cosmic neutrinos has been reported by the \ice\ Collaboration \citep{Aartsen:2013bka,Aartsen:2013jdh}; the absence of any identified point sources \citep{Aartsen:2016xlq} places firm limits on possible contributions from persistently bright neutrino sources \citep{Murase:2016gly}. 
In this context, multimessenger studies have provided important clues to the origins of the diffuse neutrino, \gray, and cosmic ray backgrounds \citep{Murase:2013rfa,Fang:2017zjf}, and offer a powerful approach for identifying transient or highly-variable neutrino sources, including blazar flares \citep{Dermer:2014vaa,Kadler:2016ygj,Petropoulou:2016ujj,turleyblazar}.
Anticipating these and related opportunities, the Astrophysical Multimessenger Observatory Network (AMON\footnote{AMON website: \url{https://www.amon.psu.edu/}}) was founded to link global HE and multimessenger observatories together into a single network and to distribute relevant alerts to the community in near real-time \citep{amon13}. 

Blazars -- active galactic nuclei oriented with a relativistic jet pointing toward Earth -- dominate the extragalactic $\gamma$-ray sky \citep{Ackermann:2014usa,TheFermi-LAT:2015ykq}. Yet, despite a wealth of electromagnetic (EM) data, blazar radiation mechanism(s) remain unclear, with leptonic and (lepto)hadronic scenarios providing viable explanations \citep[e.g.,][]{Boettcher:2013wxa}. 
Since blazars and other jetted active galactic nuclei are proposed ultrahigh-energy cosmic-ray (UHECR; $\ecr\gtrsim {\rm 3\,EeV}$) accelerators \citep[e.g.,][]{2012ApJ...749...63M}, the question of hadronic acceleration in these sources has important broader implications, and information from HE neutrinos will likely be crucial to resolving these issues \citep{Murase:2015ndr}.

The \amonehe\ alert 50579430 (hereafter, \icnu) was identified by \ice\ and publicly distributed via AMON and the Gamma-ray Coordinates Network (GCN) within $\delta t\approx 43$\,s of its interaction in the Antarctic ice cap at 20:54:30.43~UT on 2017 September~22 \citep{GCN_notice_icnu}.
As with previous likely-cosmic events, its location was soon targeted by multiple observatories covering a broad energy range. 
The \swift\ XRT \citep{gcn-xrt-1} and \fermi\ LAT \citep{atel-lat} reported an association with a blazar, \txs, which showed strong activity in LAT data beginning 2017~April, and significant \xray\ variability during \swift\ monitoring observations.  
Broadband EM observations of this event, the significance of the blazar flare, and the high-energy neutrino coincidence are discussed in \citet{Aartsen2018blazar1}.

The present work is organized as follows. 
In Sec.~\ref{sec:obs} we present a comprehensive analysis of \fermi\ (\gray), \nustar\ (hard \xray), \swift\ (\xray, ultraviolet, optical), and \xshooter\ (ultraviolet, optical, near-infrared) observations of \txs, using these data to construct the source spectral energy distribution (SED) over near-infrared to \gray\ energies for two epochs: a 30-day period centered around the time of the neutrino trigger (Ep.~1) and a 30-day period starting 15 days after the neutrino trigger (Ep.~2). 
In Sec.~\ref{sec:model} we model the SED of \txs\ in the flaring phase (Ep.~1) by performing detailed radiative transfer calculations, focusing on leptonic and hadronic single-zone emission scenarios. 
We discuss the implications of our modeling results in Sec.~\ref{sec:discuss}, and conclude in Sec.~\ref{sec:conclude}. 

Throughout the manuscript we use the notation $Q_{\rm x}$ as a shorthand for the quantity $Q/10^{\rm x}$ (with $Q$ in cgs units) unless stated otherwise. 
We note that given its redshift $z=0.3365$ \citep{Paiano:2018qeq} and a consensus cosmology, the luminosity distance of \txs\ is $d_L \approx 1750$\,Mpc. 


\section{Observations \& Analysis}
\label{sec:obs}

In this section we review the \ice\ detection of the \icnu\ neutrino, and present observations and data analysis for EM follow-up observations from the \fermi\ (\gray), \nustar\ (hard \xray), \swift\ (\xray, ultraviolet, optical), and \xshooter\ (ultraviolet, optical, near-infrared) facilities. 
We note that while very high-energy ($\egam\simgt 100$\,GeV) \gray\ observations by multiple facilities have been reported \citep{atel-hess,atel-veritas}, including a first detection at these energies by MAGIC \citep{atel-magic}, details are not yet publicly available. 
These constraints are not included in our analysis. 
Given our results, consistency of our models with \fermi\ data out to $\egam\approx 100$\,GeV, and current uncertainties in the necessary extragalactic background light corrections for the source at these energies, we do not expect that inclusion of these constraints would alter our conclusions. 


\subsection{IceCube Data}
\label{sub:obs:icecube}

\icnu\ was an EHE neutrino event \citep{GCN_AMON} identified and distributed by the \ice\ Observatory via AMON and GCN within $\delta t\approx 43$\,s of its detection at 20:54:30~UT on 2017 September 22 \citep{GCN_notice_icnu}. 
A refined localization was reported four hours later~\citep{GCN_icnu}: 
R.A.=$77.43^{+1.3}_{-0.8}$~deg, Dec.=$+5.72^{+0.7}_{-0.4}$~deg (J2000; 90\% containment ellipse). 
The maximum likelihood neutrino position is R.A. \rasxgd{05}{09}{08}{784}, 
Dec.\ \dcsxgd{+05}{45}{13}{32} (J2000); see Fig.~\ref{fig:swift-pointings} for an illustration of the initial and final localizations.

EHE neutrino event reports include the neutrino arrival time; direction (R.A. and Dec.), angular error ($r_{50}$ for 50\% containment; $r_{90}$ for 90\% containment), and revision number; an estimate of the deposited \pcharge, an estimate of the neutrino \penergy, and the parameter \psig, an estimate of the probability that the event was due to an astrophysical -- rather than atmospheric -- neutrino \citep{IC_realtime}.
Real-time identification, localization, and reporting of \ice\ HE neutrinos is enabled by software in-place at the South Pole since April~2016 \citep{IC_realtime}. 



\subsection{\swift\ XRT Data}
\label{sub:obs:swift}

\begin{figure}[th]
\begin{center}
 \includegraphics[scale=0.25]{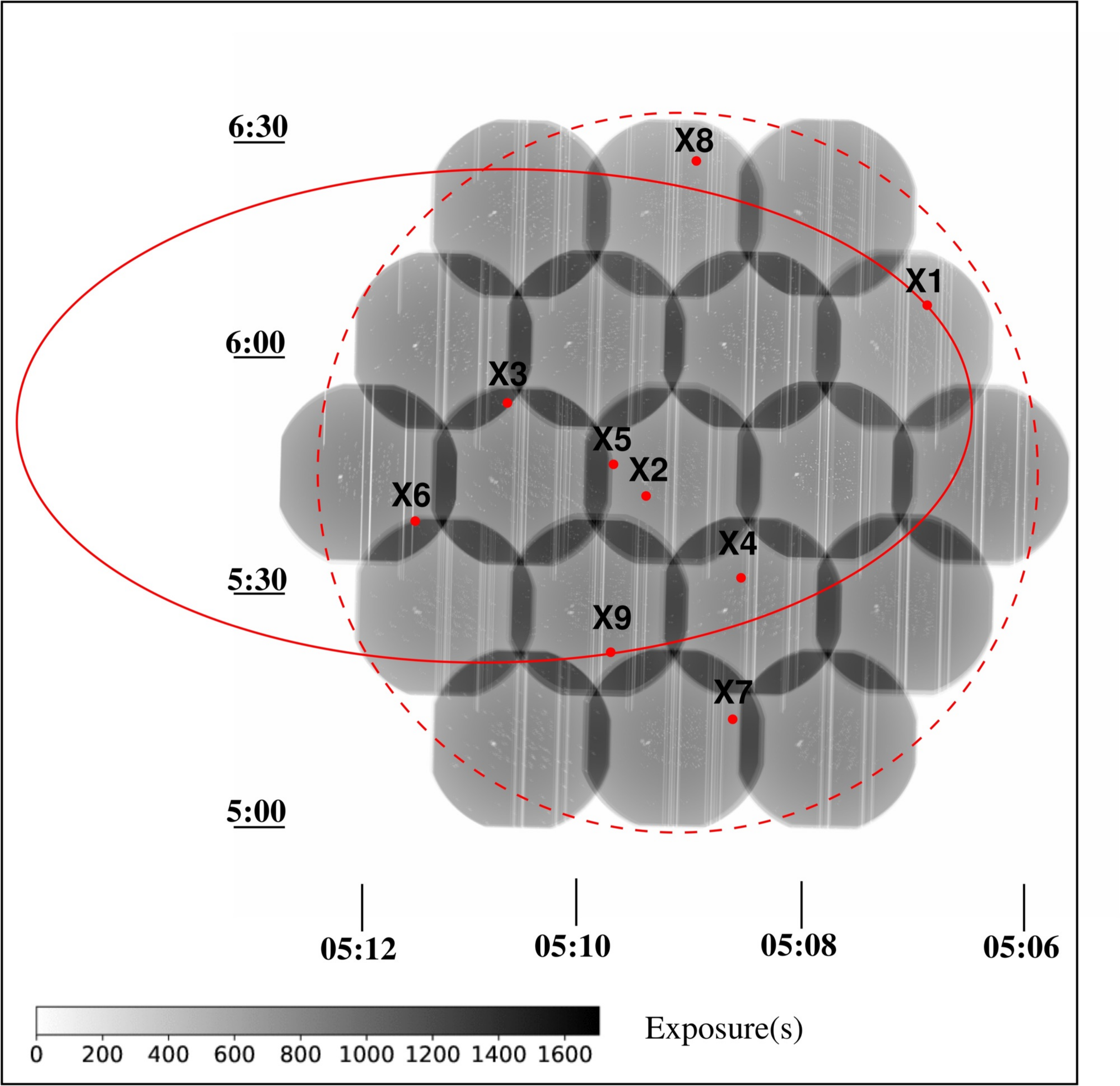}
\caption{\swift\ XRT follow up of \icnu.\ \xray\ exposure map resulting from the adopted 19-point tiling pattern centered on the initial IceCube neutrino localization is shown in gray-scale, and the positions of all detected \xray\ sources with red points.  
The red dashed circle shows the initial 90\%-containment region. 
The red solid ellipse shows the updated 90\%-containment region~\citep{GCN_icnu}.
Gray-scale levels indicate achieved exposure at each sky position, as shown by the color bar. 
White streaks are due to dead regions on the XRT detector caused by a micrometeroid impact \citep{abbey+06}.} 
\label{fig:swift-pointings}
\end{center}
\end{figure}

\icnu\ triggered \edit1{the Neil Gehrels \swift\ Observatory} in automated fashion via AMON cyberinfrastructure, resulting in rapid-response mosaic-type follow-up observations, covering a roughly circular region of sky centered on the prompt localization in a 19-point tiling that began 3.25~hours after the neutrino detection.
This initial epoch of \swift\ observations spanned 22.5~hours and accumulated $\approx$800~s exposure per pointing. 
The mosaic tiling yielded coverage of a region with radius $\approx$0.8\dg\ centered on R.A. \rasxgd{05}{09}{08}{784}, Dec.\ \dcsxgd{+05}{45}{13}{32} (J2000), amounting to a sky area of 2.1 deg$^2$.
XRT data were analyzed automatically, as data were received at the University of Leicester, via the reduction routines of~\citet{xrtauto2009,xrtcat2014}. 
Nine \xray\ sources were detected in the covered region down to a typical achieved depth of $3.8\times 10^{-13}$\,\ergcms\ (0.3--10.0 keV). 
Fig.~\ref{fig:swift-pointings} shows the exposure map for the 19-point tiling pattern, along with the nine detected \xray\ sources.
All detected sources were identified as counterparts to known and cataloged stars, \xray\ sources, or radio sources~\citep{gcn-xrt-1}; fluxes of these \xray\ sources were consistent with previously measured values.

Notably, Source~2 from these observations (marked as X2 on Fig.~\ref{fig:swift-pointings}), located 4.6\arcmin\ from the center of the neutrino localization, was identified by us as the likely \xray\ counterpart to QSO~J0509+0541, also known as \txs. This was the first report to connect \txs\ to \icnu\ \citep{gcn-xrt-1}. 

Following the \fermi\ report that \txs\ was in a rare GeV-flaring state~\citep{atel-lat}, we commenced a \swift\ monitoring campaign on September~27 \citep{atel-xrt-2}.
\swift\ monitored \txs\ for 36 epochs by November~30 with 53.7~ks total exposure time (Table~\ref{tab:xrt-monitoring}). 

To characterize the \xray\ flux and spectral variability of \txs, we performed a power-law fit to each individual \swift\ XRT observation (Table~\ref{tab:xrt-monitoring}), as well as to the summed spectrum from all listed epochs, \edit1{using XSPEC~\citep{xspec}}. 
The observation on October~14 is excluded from the spectral analysis due to low exposure time. 
The summed spectrum is adequately fit with a single power-law spectral model having the Galactic column density $N_{\rm H}=1.11\times 10^{21}$\,\percmsq, resulting in a photon index $\alpha_{\rm XRT}=2.37\pm0.05$ and mean flux of $2.27\times 10^{-12}$\,\ergcms (0.3--10.0 keV).

\begin{figure*}[th]
  \begin{center}
      \includegraphics[scale=0.29]{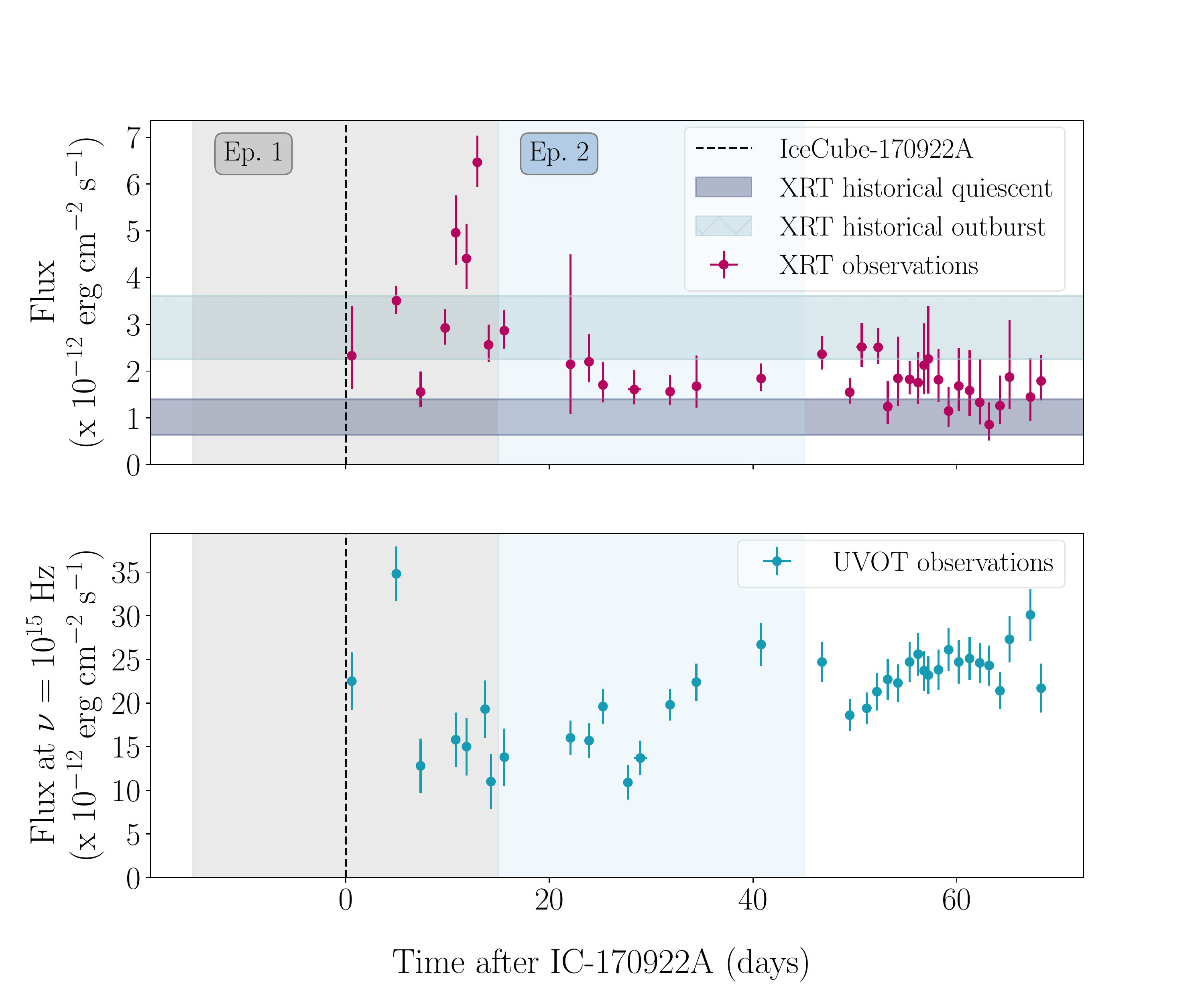}
     \includegraphics[scale=0.29]{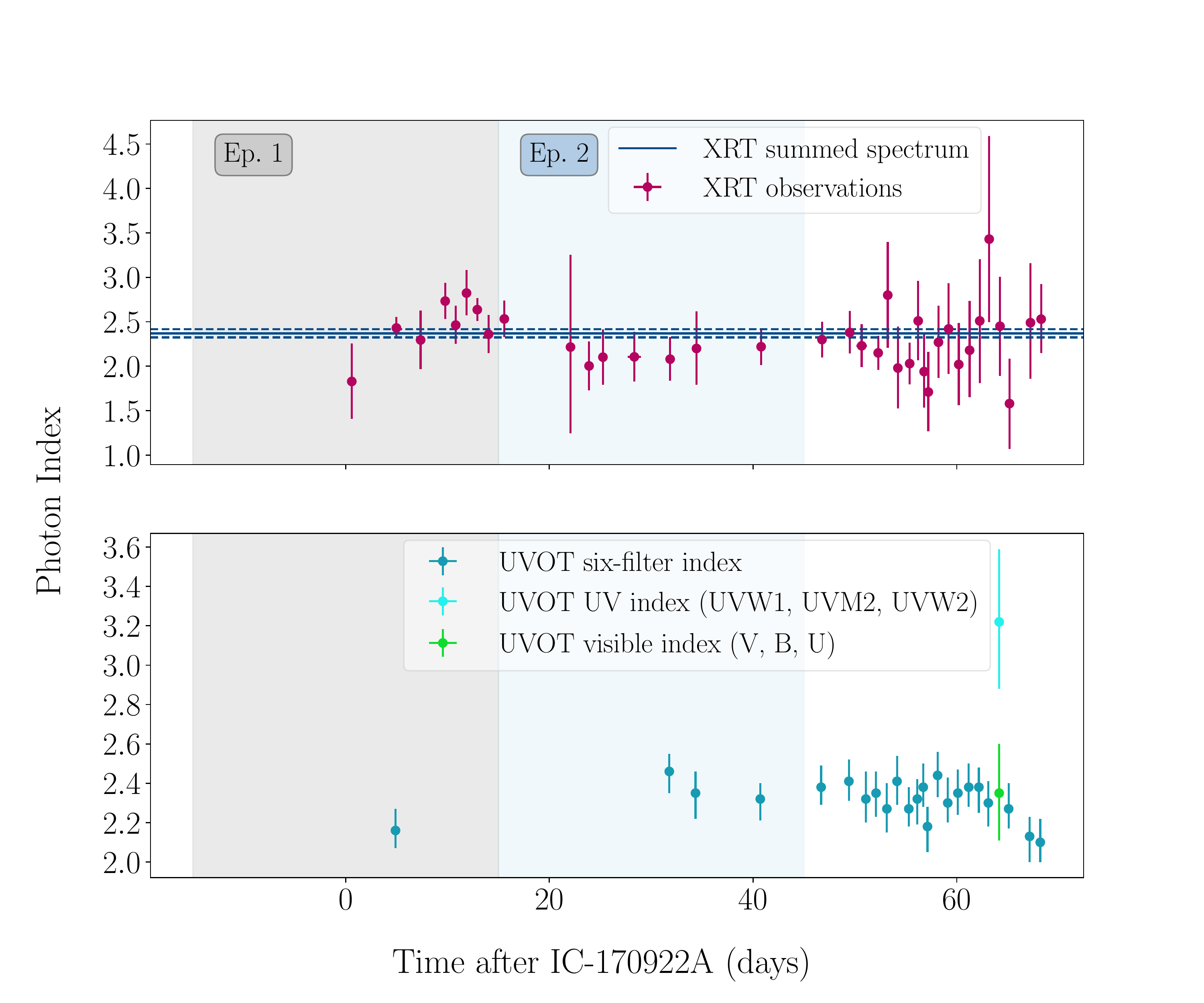}
    \caption{(Left Top) \swift\ XRT light curve. Each bin corresponds to one observation in the 0.3--10 keV energy range. The horizontal bands show the XRT historical data (four observations) of \txs:\ the mean historical quiescent flux from combining three data points, and one showing the rate from the outburst observation. 
    (Left Bottom) \swift\ UVOT light curve for all 36 observations performed on \txs.\ The dashed line shows the \icnu\ arrival time.  
    (Right top) \swift\ XRT photon index variation during the XRT monitoring campaign of \txs.\ The solid horizontal line shows the photon index of the stacked X-ray spectrum over the 2 epochs while the dashed lines represent the uncertainties. 
    (Right Bottom) \swift\ UVOT photon index variations obtained from a power-law fit to the energy flux spectrum ($\varepsilon_\gamma$ vs. $F_{\varepsilon_\gamma}$). In all plots, Ep.~1 and Ep.~2 are, respectively, defined as [-15d, +15d] and [+15d, +45d] time windows with respect to the \icnu\ arrival time.}  
\label{fig:swift-lc-index}
  \end{center}
\end{figure*}

We note that this source has been observed on multiple previous occasions with \swift\ XRT, with results published in the 1SXPS catalog~\citep{xrtcat2014}. 
In past observations, \txs\ exhibits a typical flux of $1\times 10^{-12}$\,\ergcms, with one observation at $\approx2.8\times10^{-12}$\,\ergcms\ (0.3--10.0 keV).
The source was thus in an active \xray\ flaring state by comparison to historical \xray\ measurements (Fig.~\ref{fig:swift-lc-index}, upper left). 
Photon indices and \xray\ flux measurements for each epoch are provided in Table~\ref{tab:xrt-monitoring}, and the variations in photon index are shown in Fig.~\ref{fig:swift-lc-index}~(upper right). 
The \swift\ XRT light curve is shown in Fig.~\ref{fig:swift-lc-index}~(upper left). 

\startlongtable
\begin{deluxetable*}{ccccc}
\colnumbers
\tabletypesize{\footnotesize}
\tablecaption{\swift\ XRT monitoring of \txs.
\label{tab:xrt-monitoring}}
\tablewidth{0pt}
\tablehead{
\colhead{Epoch} & \colhead{Exposure [ks]} & \colhead{Photon Index} & \colhead{$R_{{\rm X},-3}$} & \colhead{$F_{{\rm X},-12}$}}
\startdata
$ 58019.4700 \pm 0.4630$ & 0.8 & $ 1.83 ^{+ 0.43 }_{ -0.42 }$ & $ 65.8 \pm 10.1$ & $ 2.33 ^{+ 1.07 }_{ -0.72 }$ \\
$ 58023.8535 \pm 0.0660$& 4.9 & $ 2.43 \pm 0.12 $ & $ 121.0 \pm 5.3 $ & $ 3.51 ^{+ 0.32 }_{ -0.29 }$ \\
$ 58026.2273 \pm 0.0391$& 2.0 & $ 2.30 \pm 0.33 $ & $ 66.2 \pm 7.8 $ & $ 1.55 ^{+ 0.43 }_{ -0.33 }$ \\
$ 58028.6419 \pm 0.0114$& 2.0 & $ 2.73 \pm 0.20 $ & $ 117.0 \pm 8.2 $ & $ 2.92 ^{+ 0.40 }_{ -0.35 }$ \\
$ 58029.6745 \pm 0.1036$& 1.1 & $ 2.46 ^{+ 0.22 }_{ -0.21 }$ & $ 182.0 \pm 14.1 $ & $ 4.96 ^{+ 0.80 }_{ -0.70 }$ \\
$ 58030.7369 \pm 0.0385$ & 1.2 & $ 2.82 ^{+ 0.26 }_{ -0.25 }$ & $ 186.0 \pm 16.1 $ & $ 4.41 ^{+ 0.74 }_{ -0.65 }$ \\
$ 58031.7944 \pm 0.1025$ & 2.3 & $ 2.64 \pm 0.13 $ & $ 255.0 \pm 11.7 $ & $ 6.47 ^{+ 0.57 }_{ -0.53 }$ \\
$ 58032.8985 \pm 0.3428$ & 2.1 & $ 2.36 ^{+ 0.22 }_{ -0.21 }$ & $ 90.1 \pm 7.1 $ & $ 2.56 ^{+ 0.44 }_{ -0.37 }$ \\
$ 58034.4478 \pm 0.0337$ & 1.9 & $ 2.53 \pm 0.21 $ & $ 108.0 \pm 8.1 $ & $ 2.87 ^{+ 0.44 }_{ -0.38 }$ \\
$ 58040.9452 \pm 0.0010$ & 0.2 & $ 2.22 ^{+ 1.04 }_{ -0.97 }$ & $ 82.9 ^{+ 29.8 }_{ -24.1 }$ & $ 2.15 ^{+ 2.35 }_{ -1.06 }$ \\
$ 58042.7684 \pm 0.1686$ & 2.2 & $ 2.00 \pm 0.27 $ & $ 61.9 \pm 5.8 $ & $ 2.20 ^{+ 0.59 }_{ -0.44 }$ \\
$ 58044.1300 \pm 0.0696$ & 1.8 & $ 2.10 \pm 0.31 $ & $ 52.6 \pm 6.0 $ & $ 1.70 ^{+ 0.49 }_{ -0.37 }$ \\
$ 58047.2130 \pm 0.6360$ & 2.3 & $ 2.11 \pm 0.28 $ & $ 49.6 \pm 5.1 $ & $ 1.61 ^{+ 0.41 }_{ -0.32 }$ \\
$ 58050.7286 \pm 0.0446$ & 2.9 & $ 2.08 ^{+ 0.25 }_{ -0.24 }$ & $ 46.6 \pm 4.4 $ & $ 1.56 ^{+ 0.35 }_{ -0.28 }$ \\
$ 58053.3162 \pm 0.0335$ & 1.0 & $ 2.20 ^{+ 0.42 }_{ -0.41 }$ & $ 54.4 \pm 8.2 $ & $ 1.68 ^{+ 0.66 }_{ -0.46 }$ \\
$ 58059.6620 \pm 0.0729$ & 3.3 & $ 2.22 \pm 0.21 $ & $ 60.2 \pm 4.7 $ & $ 1.84 ^{+ 0.32 }_{ -0.27 }$ \\
$ 58065.6390 \pm 0.3371$ & 3.1 & $ 2.30 \pm 0.20 $ & $ 83.4 \pm 6.2 $ & $ 2.36 ^{+ 0.39 }_{ -0.33 }$ \\
$ 58068.3642 \pm 0.0742$ & 3.0 & $ 2.38 \pm 0.24 $ & $ 55.7 \pm 4.9 $ & $ 1.55 ^{+ 0.30 }_{ -0.25 }$ \\
$ 58069.5359 \pm 0.5032$ & 1.8 & $ 2.23 \pm 0.24 $ & $ 83.3 \pm 7.5 $ & $ 2.52 ^{+ 0.52 }_{ -0.42 }$ \\
$ 58071.1551 \pm 0.1299$ & 2.9 & $ 2.15 \pm 0.19 $ & $ 81.1 \pm 5.9 $ & $ 2.51 ^{+ 0.42 }_{ -0.35 }$ \\
$ 58072.0949 \pm 0.0039$ & 0.7 & $ 2.80 ^{+ 0.60 }_{ -0.59 }$ & $ 48.5 \pm 9.1 $ & $ 1.24 ^{+ 0.55 }_{ -0.37 }$ \\
$ 58073.0911 \pm 0.0042$ & 0.7 & $ 1.98 ^{+ 0.47 }_{ -0.45 }$ & $ 55.0 \pm 9.7 $ & $ 1.84 ^{+ 0.89 }_{ -0.59 }$ \\
$ 58074.2484 \pm 0.0962$ & 3.0 & $ 2.03 ^{+ 0.24 }_{ -0.23 }$ & $ 54.4 \pm 4.8 $ & $ 1.82 ^{+ 0.39 }_{ -0.32 }$ \\
$ 58075.0831 \pm 0.0046$ & 0.8 & $ 2.51 \pm 0.45 $ & $ 63.1 \pm 9.6 $ & $ 1.75 ^{+ 0.66 }_{ -0.46 }$ \\
$ 58075.6655 \pm 0.0057$ & 1.0 & $ 1.94 \pm 0.41 $ & $ 60.6 \pm 9.4 $ & $ 2.13 ^{+ 0.90 }_{ -0.62 }$ \\
$ 58076.0793 \pm 0.0049$ & 0.8 & $ 1.71 ^{+ 0.45 }_{ -0.44 }$ & $ 56.1 \pm 9.0 $ & $ 2.26 ^{+ 1.14 }_{ -0.74 }$ \\
$ 58077.0756 \pm 0.0054$ & 0.9 & $ 2.27 ^{+ 0.41 }_{ -0.40 }$ & $ 61.4 \pm 8.9 $ & $ 1.81 ^{+ 0.66 }_{ -0.48 }$ \\
$ 58078.0717 \pm 0.0056$ & 1.0 & $ 2.42 ^{+ 0.52 }_{ -0.51 }$ & $ 40.5 \pm 7.1 $ & $ 1.15 ^{+ 0.52 }_{ -0.34 }$ \\
$ 58079.0675 \pm 0.0057$ & 1.0 & $ 2.02 ^{+ 0.47 }_{ -0.46 }$ & $ 49.3 \pm 8.6 $ & $ 1.68 ^{+ 0.80 }_{ -0.53 }$ \\
$ 58080.1320 \pm 0.0057 $ & 1.0 & $ 2.18 ^{+ 0.55 }_{ -0.53 }$ & $ 47.4 \pm 9.4 $ & $ 1.58 ^{+ 0.86 }_{ -0.54 }$ \\
$ 58081.1279 \pm 0.0057$ & 1.0 & $ 2.51 ^{+ 0.69 }_{ -0.70 }$ & $ 45.8 \pm 9.7 $ & $ 1.33 ^{+ 0.93 }_{ -0.48 }$ \\
$ 58082.0551 \pm 0.0057$ & 1.0 & $ 3.43 ^{+ 1.16 }_{ -0.93 }$ & $ 30.0 ^{+ 9.4 }_{ -7.8 }$ & $ 0.85 ^{+ 0.47 }_{ -0.34 }$ \\
$ 58083.1195 \pm 0.0057$ & 1.0 & $ 2.45 \pm 0.56 $ & $ 44.4 \pm 8.0 $ & $ 1.26 ^{+ 0.65 }_{ -0.39 }$ \\
$ 58084.0499 \pm 0.0058$ & 1.0 & $ 1.58 ^{+ 0.50 }_{ -0.51 }$ & $ 39.0 \pm 6.9 $ & $ 1.87 ^{+ 1.22 }_{ -0.69 }$ \\
$ 58086.1089 \pm 0.0059$ & 1.0 & $ 2.49 ^{+ 0.67 }_{ -0.63 }$ & $ 45.4 \pm 9.8 $ & $ 1.44 ^{+ 0.84 }_{ -0.52 }$ \\
$ 58087.1560 \pm 0.0053$ & 0.9 & $ 2.53 ^{+ 0.39 }_{ -0.38 }$ & $ 66.9 \pm 9.2 $ & $ 1.79 ^{+ 0.55 }_{ -0.42 }$\\
\enddata
\tablecomments{$R_{{\rm X},-3}$ and $F_{{\rm X},-12}$ indicate count rate and energy flux, in units of $10^{-3}$~\ctsec\ and $10^{-12}$~\ergcms, respectively. Uncertainties are quoted at 90\% confidence.}
\end{deluxetable*}


\subsection{\nustar\ Data}
\label{sub:obs:nustar}
To further characterize the HE emissions of the source, we requested two observations with the \nustar\ hard \xray\ (3.0--100 keV) mission \citep{nustar2013}. 

On 2017 September~29 (02:23 to 17:48 UTC) \nustar\ carried out a Target of Opportunity observation of \txs\ \citep{atel-xrt-nustar}. 
The full science observation was retrieved from the \nustar\ public archive (ObsID 90301618002). 
Data were processed within the \edit1{{\scshape HEAsoft} \citep{xspec}} software environment using the \texttt{nupipeline} tool with the setting \texttt{SAAMODE=strict}. This yielded exposures of 23.9~ks (24.5~ks) and count rates of 21.3\,\ctksec\ (20.8\,\ctksec) in the A (B) units, respectively. 
Level~3 data products for the source were then extracted using the \texttt{nuproducts} tool.  
Within \texttt{XSPEC}, spectral data from both units were fit to a single power-law model with $N_{\rm H}=1.11\times 10^{21}$\,\percmsq, resulting in a photon index of $\alpha_{\rm NuSTAR}=1.69\pm0.12$ and a flux of $4.27^{+0.50}_{-0.58}\times 10^{-12}$\,\ergcms \ (3.0--79.0 keV). 

On 2017 October~19 (10:26 to 21:21 UTC) \nustar\ performed a second Target of Opportunity observation of \txs\ (ObsID 90301618004). 
Executing the same reduction as for the first observation yielded exposures of 19.7~ks (19.7~ks) and count rates of 20.2\,\ctksec\ (19.6\,\ctksec) in the A (B) units at this second epoch. 
Performing the same spectral fit as for the first observation, we obtain a photon index $\alpha_{\rm NuSTAR}=1.68\pm 0.14$ and flux of $3.65^{+0.54}_{-0.59}\times 10^{-12}$\,\ergcms \ (3.0--79.0 keV), consistent with results of the first observation from 20~days earlier. 


\subsection{Joint \swift\ XRT and \nustar\ Analysis}
\label{sub:obs:xrtnustar}

In order to obtain the energy spectrum for a wider \xray\ band (0.2--100 keV), we simultaneously fit data from individual XRT observations and \nustar\ for two main epochs: [-15d,+15d] (Ep.~1) and [+15d,+45d] (Ep.~2) relative to the neutrino detection.
The two epochs include eight and seven XRT observations, respectively, and one \nustar\ observation each. 
Since the source spectrum over the \nustar\ bandpass does not change from Ep.~1 to Ep.~2, we fit all individual XRT observations together with both \nustar\ observations with a sum of two power laws model, including Galactic absorption frozen at $N_{\rm H}=1.11\times 10^{21}$\,\percmsq, and quote the soft component best-fit parameters when reporting XRT results. 
A Markov Chain Monte Carlo algorithm is then employed for each fit to provide the 90\% confidence levels. 
We generate 1000 realizations of the spectra from each XRT observation and add all Ep.~1 and Ep.~2 realizations together in order to find the 90\% confidence intervals on the flux density versus energy. 
This joint analysis results in best-fit photon indices ($F_{\egam}\propto\egam^{1-\alpha}$) of $\alpha_{\rm XRT}=2.37\pm0.05$ and $\alpha_{\rm NuSTAR}=1.68\pm0.14$. 
The results are displayed in Fig.~\ref{fig:summary} and used for subsequent SED modeling (see Section~\ref{sec:model}). 
 



\subsection{\swift\ UVOT Data}
\label{sub:obs:uvot}

The Swift UltraViolet and Optical Telescope (UVOT) \citep{uvot2005} also participated in the rapid response follow-up observations of the \icnu\ and the subsequent monitoring of the flaring blazar \txs\ as described in Section~\ref{sub:obs:swift}.
The $u$ filter was used during all 19 pointings used to tile the region around \icnu.
\txs\ was readily detected during this initial survey.
All 6 UVOT lenticular filters ($v$, $b$, $u$, uvw1, uvm2, uvw2) were used during the subsequent observations monitoring \txs. 

UVOT data were analyzed using the standard tool {\scshape uvotsource} of {\scshape HEAsoft} \edit1{(v6.22.1)} and the latest updates to the UVOT CALDB files.
{\scshape uvotsource} does aperture photometry \citep{breeveld2010,breeveld2011} using user-specified source and background regions.
Because \txs\ is a bright source, a 5\arcsec\ aperture was used for the source region.
A nearby source-free region was used as the background region instead of the usual concentric ring centered on the target to avoid contamination from a read-out streak.
The read-out streak is produced by photons from a very bright source near the edge of the UVOT image that arrive during the brief time in which the image frame is transferred for read out of the detector's CCD \citep{page2013}.
The position of the read-out streak changes from observation to observation as the orientation on the sky of the UVOT image changes.
Consequently a different nearby background region was used for the observations centered near MJD 58065 (Swift sequence 00083368018). 
Data from observations near MJD 58028.6 (sequence 00083368003) and MJD 58031.8 (sequence 00083368006) are not used because the source region is within the read-out streak.

Table~\ref{tab:uvot} reports the times, exposures, and magnitudes for all the observations. 
The source varies over a range of at least 0.5 magnitudes in all 6 filters. 

\startlongtable
\begin{deluxetable}{cccc}
\colnumbers
\tabletypesize{\footnotesize}
\tablecaption{\swift\ UVOT monitoring of \txs 
\label{tab:uvot}}
\tablewidth{0pt}
\tablehead{
\colhead{Epoch$^a$ (MJD)} &
\colhead{Exposure (s)} &
\colhead{Filter} &
\colhead{Magnitude$^b$}}
\startdata
$ 58019.4699 \pm 0.4632$ &  780 & u & $14.31 \pm 0.03$\\
$ 58023.8555 \pm 0.0641$ &  809 & uvw2 & $14.58 \pm 0.03$\\
$ 58023.8257 \pm 0.0306$ &  157 & v & $14.62 \pm 0.04$\\
$ 58023.8364 \pm 0.0403$ & 2930 & uvm2 & $14.50 \pm 0.03$\\
$ 58023.8515 \pm 0.0639$ &  472 & uvw1 & $14.36 \pm 0.04$\\
$ 58023.8529 \pm 0.0635$ &  236 & u & $14.27 \pm 0.04$\\
$ 58023.8539 \pm 0.0635$ &  236 & b & $15.08 \pm 0.03$\\
$ 58026.2274 \pm 0.0391$ & 1954 & uvm2 & $14.81 \pm 0.04$\\
\enddata
\tablecomments{Table~\ref{tab:uvot} is published in its entirety in the machine-readable format.
A portion is shown here for guidance regarding its form and content.}
\tablenotetext{a}{MJD at the middle of the observation.}
\tablenotetext{b}{Errors at 1-$\sigma$ uncertainty.}
\end{deluxetable}


\subsection{\swift\ UVOT Analysis}
\label{sub:obs:uvotsed}

Due to the UVOT's blue response, extending as far as $\lambda \approx 1600$\AA, it is necessary to apply an appropriate extinction correction in order to interpret UVOT observations appropriately and derive physical constraints on the SED of \txs. 
The line of sight extinction to \txs\ as provided by the NASA/IPAC Infrared Science Archive\footnote{NASA/IPAC Infrared Science Archive: \url{https://irsa.ipac.caltech.edu/}} is $A_V = 0.286$\,mag according to all-sky dust maps~\citep{sfd98,sf11}. 
Although this estimate nominally has little uncertainty ($\delta A_V=0.008$\,mag), we note that (1) This quoted value has been corrected from the original published value by 14\% \citep{sf11}; and (2) A subsequent recalibration with a similar approach, using \mbox{PanSTARRS-1} rather than Sloan Digital Sky Survey data \citep{gsf+15}, leads to a different extinction estimate, $A_V =0.254$\,mag. 
We therefore take a conservative approach and adopt a line of sight extinction of $A_V = 0.286\pm 0.032$\,mag, with Gaussian form, as our prior for UVOT SED analysis. 
To calculate extinction as a function of wavelength we use the \citet{Fitzpatrick1999} extinction law  with $R_V=3.1$. 

On one occasion during Ep.~1, on three occasions during Ep.~2, and on twenty subsequent occasions before November~30, UVOT observations were carried out using all six UV/optical filters in rapid sequence. 
We use these observations to characterize the UV/optical SED and UVOT spectral index (Fig.~\ref{fig:swift-lc-index}, lower right). 
Additional single-filter observations are then used (in tandem with the six-filter epochs) to estimate source variability (Fig.~\ref{fig:swift-lc-index}, lower left). 

To fit the SED for each six-filter UVOT observation, we perform a $\chi^2$ minimization of the predicted versus observed count rates in each filter using a model with three parameters:  source flux density $F_\nu$ at $\nu=\egam/h=10^{15}$\,Hz, UVOT spectral index $\beta$, and extinction $A_V$. 
Extinction values away from our adopted value of $A_V=0.286$\,mag are penalized according to our Gaussian extinction uncertainty. 
The source spectrum is integrated across each filter bandpass using the filter transmission function\footnote{UVOT Calibrations Database (CALDB):\\ \url{https://heasarc.gsfc.nasa.gov/docs/heasarc/caldb/data/swift/uvota/index.html}} and \citet{Fitzpatrick1999} extinction curve. 
A fit is considered successful if the total $\chi^2$ across the six filters gives a $p$-value $p(\chi^2>\chi^2_{\rm obs}) > 5\%$. 
Only one of the six-filter observations (at $\delta t = 64.5$\,days after \icnu) violates this constraint, and so receives treatment as a broken power-law across the UVOT filters \edit1{(forcing a single power-law fit gives a spectral index $\beta=1.66\pm 0.06$ at this epoch, where we quote uncertainties from $\delta\chi^2$ analysis even though the fit is acknowledged not to be satisfactory)}. 
We note that curvature of the source spectrum (in particular, steepening/softening toward the UV) is also observed in the \xshooter\ data (Sec.~\ref{sub:obs:xshooter}). 
Acceptable fits yield a best-fit value for the UVOT spectral index and uncertainty via $\Delta\chi^2$ analysis. 
These values and uncertainties are reported in Table~\ref{tab:uvot-index} and plotted in Fig.~\ref{fig:swift-lc-index} (lower right). 


\begin{table}[bt]
\caption{\swift\ UVOT photon indices for \txs.}
\label{tab:uvot-index}
\begin{center}
\begin{tabular}[t]{lc}
\hline
Epoch & Photon Index \\ \hline 
58023.7632 & $ 2.16 ^{+ 0.11 }_{- 0.09 }$ \\ 
58050.6452 & $ 2.46 ^{+ 0.09 }_{- 0.11 }$ \\ 
58053.2192 & $ 2.35 ^{+ 0.11 }_{- 0.13 }$ \\ 
58059.5782 & $ 2.32 ^{+ 0.08 }_{- 0.11 }$ \\ 
58065.5562 & $ 2.38 ^{+ 0.11 }_{- 0.09 }$ \\ 
58068.2802 & $ 2.41 ^{+ 0.11 }_{- 0.10 }$ \\ 
58069.9502 & $ 2.32 ^{+ 0.14 }_{- 0.12 }$ \\ 
58070.9472 & $ 2.35 ^{+ 0.11 }_{- 0.12 }$ \\ 
58072.0112 & $ 2.27 ^{+ 0.13 }_{- 0.12 }$ \\ 
58073.0082 & $ 2.41 ^{+ 0.13 }_{- 0.12 }$ \\ 
58074.1632 & $ 2.27 ^{+ 0.11 }_{- 0.09 }$ \\ 
58074.9992 & $ 2.32 ^{+ 0.10 }_{-0.13 }$ \\ 
58075.5822 & $ 2.38 ^{+ 0.12 }_{-0.10 }$ \\ 
58075.9962 & $ 2.18 ^{+ 0.10 }_{-0.13 }$ \\ 
58076.9922 & $ 2.44 ^{+ 0.12 }_{-0.11 }$ \\ 
58077.9882 & $ 2.30 ^{+ 0.13 }_{-0.10 }$ \\ 
58078.9842 & $ 2.35 ^{+ 0.12 }_{-0.11 }$ \\ 
58080.0482 & $ 2.38 ^{+ 0.12 }_{-0.10 }$ \\ 
58081.0442 & $ 2.38 ^{+ 0.10 }_{-0.13 }$ \\ 
58081.9712 & $ 2.30 ^{+ 0.11 }_{-0.12 }$ \\ 
58083.9662 & $ 2.27 ^{+ 0.13 }_{-0.10 }$ \\ 
58086.0252 & $ 2.13 ^{+ 0.10 }_{-0.13 }$ \\ 
58083.0362 (UV) & $ 3.22 ^{+ 0.37 }_{-0.34 }$ \\ 
58083.0362 (Vis) & $ 2.35 ^{+ 0.25 }_{-0.24 }$ \\ 
58087.0722 & $ 2.10 ^{+ 0.12 }_{-0.10 }$ \\ \hline
\end{tabular}\\
\end{center}
\vspace{0.5\baselineskip}
\tablecomments{UVOT photon indices from power-law SED fitting for epochs with data in all 6 filters. Data for the next-to-last epoch are fitted with a power-law to the 3 UV and 3 visible filters separately; \edit1{a forced single power-law fit to this epoch yields a photon index of 2.66,} see text for details.}
\end{table}

We determine UVOT SED bands, the range of fluxes allowed at 90\% confidence for each photon energy, by drawing 1000 samples according to the $\chi^2$ probability function, and generating a power-law spectrum across the UVOT bandpass for each. 
The allowed range at each energy in the SED is defined as the minimum-width range encompassing 90\% of these spectra. 
For Ep.~2 we draw 1000 samples from each of the three six-filter epochs and combine these before finding the 90\%-confidence range; results for this epoch thus account for source flux variability. 
For Ep.~1 we use only the single six-filter observation to characterize the SED; we estimate accounting for flux variability over Ep.~1 would expand this band by 12\%; however, we do not make this correction in our analysis. 

We determine UVOT fluxes for single-epoch observations using the $\beta$ measurement from the temporally proximate six-filter observation, adjusting the flux density at $10^{15}$\,Hz to achieve agreement with the observed count rate in the relevant filter. 
Quoted uncertainties for these flux estimates combine the Poisson count rate uncertainty with the uncertainty in flux for the adopted SED model, in quadrature. 
Flux values and uncertainties are reported in Table~\ref{tab:uvot-fluxes} and plotted in Fig.~\ref{fig:swift-lc-index} (left). 


\begin{table}
\caption{\swift\ UVOT extinction-corrected fluxes at $\nu=10^{15}$\,Hz}
\label{tab:uvot-fluxes}
\begin{center}
\begin{tabular}{c c}
\hline
Epoch & $F_{{\rm UV},-12}$ \\
\hline
58019.4709 & $ 22.50 \pm 3.29 $ \\ 
58023.8470 & $ 34.80 \pm 3.13 $ \\ 
58026.2284 & $ 12.80 \pm 3.13 $ \\ 
58029.6756 & $ 15.80 \pm 3.13 $ \\ 
58030.7380 & $ 15.00 \pm 3.29 $ \\ 
58032.5593 & $ 19.30 \pm 3.29 $ \\ 
58033.1319 & $ 11.00 \pm 3.13 $ \\ 
58034.4489 & $ 13.80 \pm 3.29 $ \\ 
58040.9463 & $ 16.00 \pm 1.97 $ \\ 
58042.7696 & $ 15.70 \pm 1.97 $ \\ 
58044.1311 & $ 19.60 \pm 1.97 $ \\ 
58046.5814 & $ 10.90 \pm 1.97 $ \\ 
58047.8132 & $ 13.70 \pm 1.97 $ \\ 
58050.7285 & $ 19.80 \pm 1.81 $ \\ 
58053.3030 & $ 22.40 \pm 2.14 $ \\ 
58059.6613 & $ 26.70 \pm 2.47 $ \\ 
58065.6395 & $ 24.70 \pm 2.30 $ \\ 
58068.3633 & $ 18.60 \pm 1.81 $ \\ 
58070.0338 & $ 19.40 \pm 1.81 $ \\ 
58071.0301 & $ 21.30 \pm 2.14 $ \\ 
58072.0950 & $ 22.70 \pm 2.30 $ \\ 
58073.0911 & $ 22.30 \pm 2.14 $ \\ 
58074.2466 & $ 24.70 \pm 2.30 $ \\ 
58075.0830 & $ 25.60 \pm 2.47 $ \\ 
58075.6651 & $ 23.70 \pm 2.30 $ \\ 
58076.0792 & $ 23.20 \pm 2.14 $ \\ 
58077.0751 & $ 23.80 \pm 2.30 $ \\ 
58078.0713 & $ 26.10 \pm 2.47 $ \\ 
58079.0670 & $ 24.70 \pm 2.47 $ \\ 
58080.1316 & $ 25.10 \pm 2.47 $ \\ 
58081.1274 & $ 24.60 \pm 2.30 $ \\ 
58082.0546 & $ 24.30 \pm 2.30 $ \\ 
58083.1191 & $ 21.40 \pm 2.14 $ \\ 
58084.0496 & $ 27.30 \pm 2.63 $ \\ 
58086.1085 & $ 30.10 \pm 2.96 $ \\ 
58087.1557 & $ 21.70 \pm 2.80 $ \\ \hline
\end{tabular}\\
\end{center}
\vspace{0.5\baselineskip}
\tablecomments{$F_{{\rm UV},-12}$ is the $\nu F_\nu(=\eFegam)$ flux at $\nu=10^{15}$\,Hz in units of $10^{-12}$\,\ergcms, derived by SED fitting for six-filter epochs, and adjusted to $\nu=10^{15}$\,Hz using the nearest best-fit SED index for single-filter epochs.}
\end{table}


\subsection{\xshooter\ Data}
\label{sub:obs:xshooter}

Medium-resolution spectroscopy of \txs\ was obtained with the \xshooter\ spectrograph \citep{Vernet2011} mounted on the Very Large Telescope UT2 at ESO Paranal Observatory on 2017 October~1. 
The three arms of \xshooter,\ (UV: UVB, optical: VIS and near-infrared: NIR) were used with slit widths of 1\farcs 0, 0\farcs 9, and 0\farcs 9, respectively. 
These data provide simultaneous 300--2480 nm spectral coverage with average spectral resolutions $\lambda/\Delta\lambda$ of 4290, 7410, and 5410, respectively, in each arm. 
Observing conditions were good, with a clear sky, seeing of $\sim$0.8\arcsec, and an airmass ranging from 1.2 to 1.3. 
Individual exposure times are 72~s, 139~s, and 54~s for the UBV, VIS, and NIR arms, respectively, and sum to total integration times of 1152~s, 2224~s, and 864~s. 
Standard ABBA nodding observing mode was used to allow for an effective background subtraction. 

Data were reduced using the ESO \xshooter\ pipeline \citep{Goldoni2006, Modigliani2010} (v.2.9.3) in the \texttt{Reflex} environment \citep{Freudling2013}, producing a background-subtracted, wavelength-calibrated spectrum. 
The extracted 1D spectrum was flux calibrated with the \xshooter\ pipeline using a response function produced by observing the \textit{HST} white dwarf standard GD71 (R.A. \rasxg{05}{52}{27}{.86}, Dec.\ \dcsxg{+15}{53}{13}{.8}, J2000) just after the observation of \txs. 
To correct results for slit losses, the final spectrum was rescaled to match the broadband B, V and R magnitudes obtained on 2017 October~29 using the 1-meter Kapteyn Telescope at La Palma \citep{Keel2017}. 
Overall, the flux calibration is expected to be accurate to 10\% in the UVB arm and 15\% in both the VIS and NIR arms based on the seeing conditions at the observing time. 
The telluric absorption lines were removed by using the \texttt{Molecfit} software through a fit of synthetic transmission spectra calculated by a radiative transfer code~\citep{Smette2015,Kausch2015}. 
Finally, we corrected the spectra for Galactic extinction using the extinction law of \citep{Fitzpatrick1999} with a total extinction at the V filter band $A_V=0.286$\,mag and a selective-to-total extinction ratio equal to the Galactic average value $R_V=3.1$. 
For the 90\%-confidence band plotted in Fig.~\ref{fig:summary}, we allowed the extinction to vary by $\delta A_V = \pm$0.054\,mag, according to our adopted uncertainty (Sec.~\ref{sub:obs:uvotsed}). 

Two Galactic interstellar absorption features are observed in the reduced spectrum: Ca~K \& H absorption lines (at 3933.7~\AA\ and 3968.~\AA\ respectively), and the Na~ID doublet at 5892.5~\AA. No other emission or absorption line is observed. 
Overall, the spectrum is consistent with the spectrum of a non-thermally dominated blazar, and confirms the source to exhibit a bluer spectrum than published by \citet{Halpern2003}, as previously mentioned by \citet{Steele2017}.  

Both the \xshooter\ and \swift\ UVOT data clearly show that the synchrotron peak is below $3\times 10^{14}$~Hz. 
This indicates that \txs\ is an intermediate synchrotron peaked (ISP) or low synchrotron peaked (LSP) blazar.   

The non-detection of Lyman-alpha absorption in the \xshooter\ spectrum \edit1{provides a rough upper limit on the redshift of \txs, $z<1.6$, which is compatible with the redshift measurement ($z=0.3365\pm 0.0010$) of \citet{Paiano:2018qeq}. }


\subsection{\fermi\ Data}
\label{sub:obs:fermi}

The \fermi\ Large Area Telescope (LAT) is a pair conversion telescope sensitive to $\gamma$ rays in the 20 MeV to $>$300 GeV~\citep{Atwood:2009ez}. 
In this section we analyze photons detected by the LAT during our defined Ep.~1 ($\pm 15$ days from the neutrino detection) and Ep.~2 (15 to 45 days after the neutrino detection). 
Analysis was performed using version \texttt{v10r0p5} of the \fermi\ \texttt{Science Tools}\footnote{\fermi\ \texttt{Science Tools} can be downloaded from \url{https://fermi.gsfc.nasa.gov/ssc/data/analysis/software/}}. 

Photons with energies between 100~MeV and 300~GeV that were detected within a radius of $15^{\circ}$ from the location of \txs\ were selected for the analysis, while photons with a zenith angle $>90^{\circ}$ were discarded to reduce contamination from the Earth's albedo. 

\begin{figure*}
\begin{center}
\includegraphics[scale=0.13]{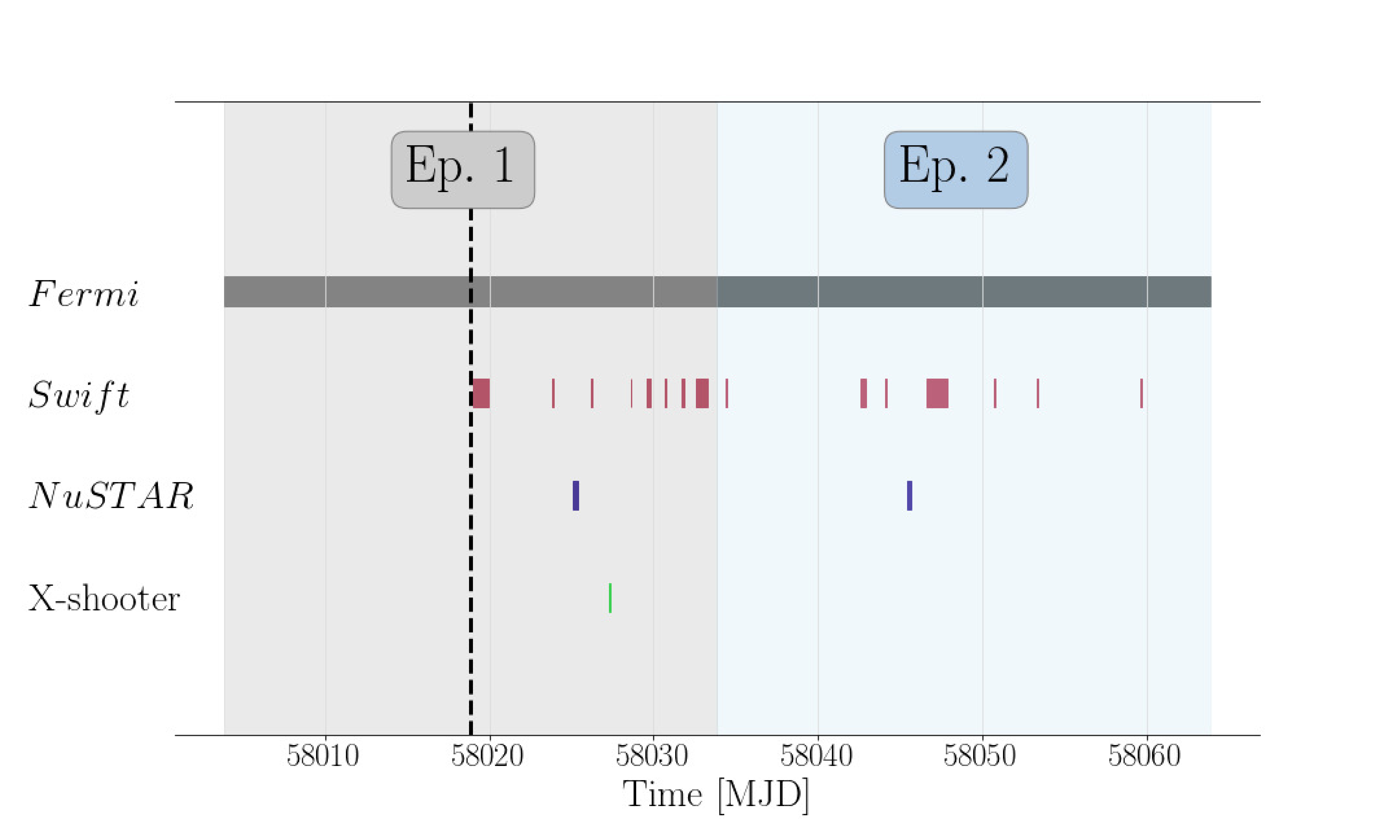}
\includegraphics[scale=0.30]{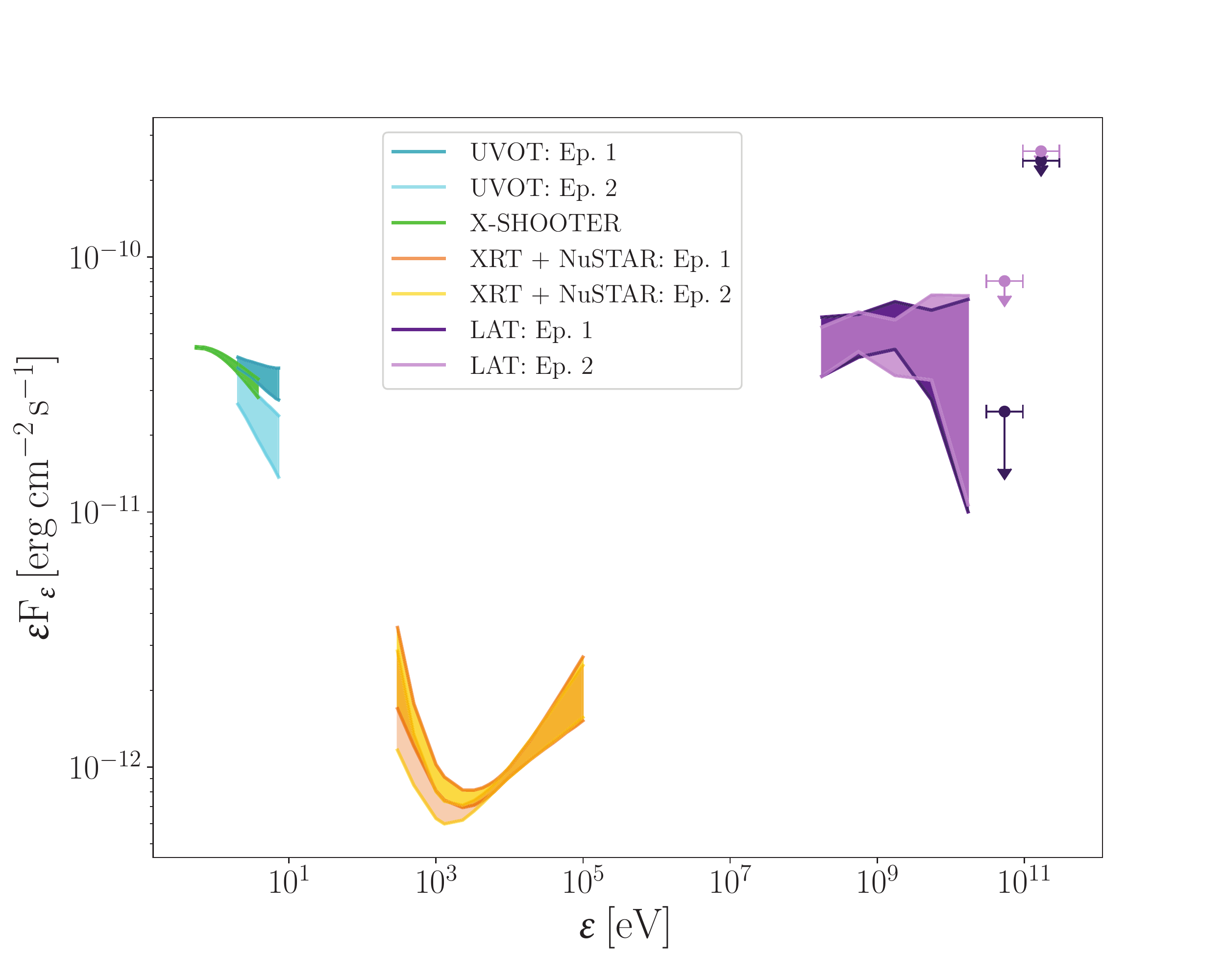}
\caption{Observations and spectral energy distribution (SED) for \txs\ in its high-flux state. 
Left panel: Timeline of observations by \fermi, \swift, \nustar, and \xshooter. Observations are divided into two 30-day epochs each for analysis and discussion purposes; the vertical dashed line shows the \icnu\ detection time. 
Right panel: Multi-wavelength SED for \txs; data with the 90\%-confidence bands on source emission are shown separately for the two epochs for each facility. The SEDs for Ep.~1 and Ep.~2 are broadly similar, with the source fading somewhat at optical through \xray\ energies, and the ultraviolet/optical spectrum softening.}  
    \label{fig:summary}
  \end{center}
\end{figure*}

The contribution from isotropic and Galactic diffuse backgrounds was modeled using the parametrization provided in the files \texttt{iso\_P8R2\_SOURCE\_V6\_v06.txt} and \texttt{gll\_iem\_v06.fits}, respectively. 
Sources in the 3FGL catalog within a radius of $15^{\circ}$ from the source position were included in the model, with spectral parameters fixed to their catalog values, while spectral parameters for sources within $3^{\circ}$ were allowed to vary freely during the fit. 
The \txs\ spectral fit was performed with a binned likelihood method using the \texttt{P8R2\_SOURCE\_V6} instrument response function.   
A power-law fit to the spectrum gives a photon index of $\alpha_{\mathrm{LAT}}=2.05\pm0.05$, consistent with the 3FGL value of $2.04\pm0.03$, and a flux normalization of $(1.42\pm 0.11)\times10^{-11}$ cm$^{-2}$ s$^{-1}$ MeV$^{-1}$ at an energy of 1.44 GeV, about a factor of four higher than the 3FGL value of $(3.24\pm0.10)\times10^{-12}$ in the same units. 
The spectral fit was repeated in seven independent energy bins with equal logarithmic spacing in the 100 MeV--300 GeV range to be incorporated in the modeling of the SED. 
Best-fit flux values and 90\% uncertainties, shown in Fig.~\ref{fig:summary} and subsequent figures, are reported for spectral bins with a test statistic (TS) value larger than 9, which corresponds to an excess of $\sim3\sigma$. 
Flux upper limits at 95\% confidence level are quoted otherwise. 





\section{Multimessenger Modeling}
\label{sec:model}

Traditionally, blazar SEDs are interpreted in two different ways. In the leptonic scenario, the $\gamma$-ray component is interpreted as synchrotron self-Compton (SSC) emission or external inverse-Compton (EIC) emission~\citep[e.g.,][]{maraschietal92,1993ApJ...416..458D,sikora94}. 
In the SSC model, the seed photons for Compton scattering are produced internally in the blazar jet; in particular, these are the synchrotron photons produced by non-thermal electron-positron pairs accelerated in the jet. 
In the EIC model, the seeds for Compton scattering are provided by external radiation fields, such as scattered accretion disk radiation, broadline/dust emission, and soft radiation from the sheath region of a structured jet. 

It is natural that protons and nuclei are also accelerated in the jet, leading to the so-called leptohadronic scenario\footnote{We will refer to this scenario simply as hadronic for simplicity.} where the $\gamma$-ray emission is explained by processes related to relativistic protons: proton-induced electromagnetic cascades~\citep{mannheim93,muecke03}, proton synchrotron emission~\citep{aharonian00,muecke03}, or intergalactic magnetic cascades induced by UHECRs~\citep{2011ApJ...731...51E,2012ApJ...749...63M}. 
In the presence of relativistic protons, theory has predicted that PeV--EeV neutrinos can be produced via the photomeson production process between cosmic-ray protons and target photons provided by the intra-jet and/or external radiation fields~\citep[see][for a recent review on AGN neutrinos and references therein]{Murase:2015ndr}. 
For example, a neutrino with $\enu\approx 0.1$~PeV to $1$~PeV implies a parent proton with energy  $\epro\approx 2.0$~PeV to $20$~PeV, for which photomeson production mainly occurs with target photons with UV \edit1{or greater} energies.



HE neutrinos generated by photohadronic interactions must be accompanied by EM emission of secondary electron-positron pairs and pionic $\gamma$ rays. 
EM cascades redistribute energy from high energies (e.g.\ PeV) to lower energies (e.g.\ keV--MeV) and exhibit $F_\gamma\sim F_\nu$. 
These cascade effects are included in our detailed numerical calculations, as \edit1{presented} in the following sections. 


\subsection{Model description}
\label{sec:setup}


We assume that protons and electrons are co-accelerated by some mechanism, whose details lie outside the immediate scope of this work, and are subsequently injected isotropically in a spherical region containing a tangled magnetic field. 
The particle interactions with the magnetic field and with secondary particles leads to the development of a system with five stable particle populations \edit1{in steady state:} protons, which lose energy by synchrotron radiation, Bethe-Heitler pair production, and photomeson production processes; electrons and positrons, which lose energy by synchrotron radiation and IC scattering; photons, which gain and lose energy in a variety of ways; neutrons, which can escape almost unimpeded from the source region, with a certain probability of photohadronic interactions; and neutrinos, which escape without any attenuation. 
The interplay of the processes governing the evolution of the energy distributions of those five populations is formulated with a set of time-dependent kinetic equations. 
Through them, energy is conserved in a self-consistent manner, since all the energy gained by a particle type has to come from an equal amount of energy lost by another particle type. 

To simultaneously solve the coupled kinetic equations for all particle types we use the time-dependent code described in \citep{DMPR12}. 
Photomeson production processes are modeled using the results of the Monte Carlo event generator {\sc SOPHIA}~\citep{SOPHIA2000}, while the Bethe-Heitler pair production is similarly modeled with the Monte Carlo results of \citet{Protheroe1996} and \citet{mastetal05}. 
The only particles that are not modeled with kinetic equations are muons, pions, and kaons~\citep{DPM14,petroetal14}; their energy losses can be safely ignored for the parameter values relevant to this study \edit1{(see also \citealt{Murase:2014foa} for numerical calculations where the kinetic equations for these particles are explicitly solved).}
 
The parameters that describe the source (i.e., Doppler factor $\delta$, comoving magnetic field strength $B^\prime$, and comoving blob size $R^\prime$) as well as these of accelerated (i.e., primary) particle distributions can often be constrained by multi-wavelength data  \citep{1996ApJ...470L..89T,mastkirk97,1997ApJ...475...97R,1998A&A...333..452K,2000ApJ...536..729L}; \edit1{a complete list of model parameters is provided in Table~\ref{tab:parameters}}. 

We search for models that \edit1{adequately} describe the multi-wavelength data (i.e., the model curve passes through most of the \edit1{instrument-specific SED bands in Fig.~\ref{fig:summary}}). 
We begin the parameter space search \edit1{using} values that we obtain analytically from expressions that relate observables to model parameters, \edit1{as described in} \citet{2012ApJ...749...63M} and \citet{Petropoulou:2015upa}. 
As we do not perform a statistical fit to the whole multi-wavelength data in the strict sense (i.e., by maximizing a likelihood function), no uncertainty ranges for the model parameters can be formally computed. 
However, thanks to the detailed quasi-simultaneous \xray\ data obtained in this work, we can place limits on the HE neutrino flux without depending on details of the model uncertainties (see subsequent sections). 
Quantitative upper limits on the proton and neutrino luminosities are placed by the requirement that the EM cascade does not \edit1{overproduce emission in the \xray\ regime (0.1--100 keV), where the source SED exhibits a prominent dip.}

The resulting upper limits are quite robust, as they depend on the energy flux ratio of the EM and neutrino components \edit1{-- determined by well-known particle interactions -- as well as the properties of EM cascades, which reliably yield a flat, broadband component} by redistributing energy from high to low energies. 

\begin{table*}[b]
\caption{Physical parameters (description, symbol, and units) used in blazar leptonic and hadronic modeling}
\begin{center}
\begin{tabular}{ccc}
\hline 
Parameter & Symbol & Unit [in cgs]\\
\hline 
Doppler factor & $\delta$ & n/a \\
Magnetic field strength & $B^\prime$ & G \\
Blob radius & $R^\prime$ & cm \\
Electron injection luminosity & $L_e^\prime$ & erg s$^{-1}$ \\
Minimum electron Lorentz factor & $\gamma^\prime_{e, \min}$ & n/a \\
Maximum electron Lorentz factor & $\gamma^\prime_{e, \max}$ & n/a \\
Break electron Lorentz factor & $\gamma^\prime_{e, br}$ & n/a \\
Power-law electron index below the break &  $s_{e,1}$ & n/a \\
Power-law electron index above the break &  $s_{e,2}$ & n/a \\
Proton injection luminosity & $L_p^\prime$ & erg s$^{-1}$ \\
Minimum proton Lorentz factor & $\gamma^\prime_{p, \min}$ & n/a \\
Maximum  proton Lorentz factor & $\gamma^\prime_{p, \max}$ & n/a \\
Power-law proton index &  $s_{p}$ & n/a \\ 
Energy density of external radiation & $u^\prime_{\rm ext}$ & erg cm$^{-3}$ \\
Effective temperature of black-body external radiation & $T^\prime$ & K \\
Photon index of power-law external  radiation & $\alpha$ &  n/a \\
Minimum photon energy of power-law external  radiation & $\epsilon^\prime_{\min}$ &  keV \\
Maximum photon energy of power-law external  radiation & $\epsilon^\prime_{\max}$ &  keV \\
\hline
\end{tabular}
\end{center}
\tablecomments{Primed quantities are measured in the jet comoving frame. Parameters describing the relativistic particle distributions refer to their properties at injection.}
\label{tab:parameters}
\end{table*}

\subsection{Leptonic Models (LMs)}
\label{sec:leptonic}

In the leptonic scenario, the blazar's SED (optical to $\gamma$-rays) is explained by synchrotron and IC processes of accelerated (primary) electrons~\citep{maraschietal92,1993ApJ...416..458D,sikora94}. 
The radiation produced by relativistic protons in the source, which are necessary for the production of HE neutrinos, may not be directly observed due to the two-photon annihilation process and subsequent EM cascades inside the source. 
We coin these hybrid scenarios ``LMs'', which stand for {\sl Leptonic Models}, \edit1{in reference to} the leptonic origin of the \grays.
Significant intra-source \gray\ attenuation at sufficiently high energies and the associated EM cascade is unavoidable in single-zone models, because target photons responsible for photohadronic interactions hinder HE $\gamma$ rays from leaving the source.
This implies that a source with efficient HE neutrino production can be \gray\ dark and may even be regarded as a hidden cosmic-ray accelerator~\citep{murase16}. 

The photomeson production process also leads to the production of \gray\ photons from neutral pion decay. 
Moreover, the decay of charged pions leads to the production of secondary electrons and positrons, which also emit HE photons via synchrotron and IC processes. 
The HE photons can be attenuated by low-energy photons in the source, while enhancing the number of secondary electron-positron pairs. 
The total absorbed photon luminosity will eventually be redistributed at lower photon energies through the development of an EM cascade~\citep{mannheim91,mannheim93}. 

The IC \edit1{emission} of primary electrons explains the HE peak of the SED, and the emission from the EM cascade should be subdominant. 
We can therefore set an upper limit on the power of the cosmic-ray proton component by requiring that any proton-induced emission does not \edit1{fill in the dip} (in hard \xrays\ for ISPs, as here) between the two peaks of the SED. 
In turn, this translates into an {\sl upper limit} on the blazar's neutrino flux.

\edit1{We first derive the maximum neutrino flux expected in the leptonic scenario by assuming that the proton distribution is a power-law with a proton index of $s_{p}=2$, extending from $\gamma'_{p, \min}=1$ to $\gamma'_{p,\max}=1.6\times{10}^7$. From the \xray\ and \gray\ light curves we infer a variability time scale of $t_{\rm var}\lesssim{10}^5$~s. Our choice of $R'=10^{17}$~cm is broadly consistent with the size inferred from the variability, namely $R'\approx\delta ct_{\rm var}/(1+z)\simeq0.56\times{10}^{17}(\delta/25)(t_{\rm var}/10^5~\rm s)$~cm. We also consider an arbitrary external photon field with a black-body--like energy distribution that can be described by only two free parameters: its characteristic temperature $T^\prime$ and energy density $u_{\rm ext}^\prime$, as measured in the comoving frame of the source. \edit1{We also neglect any angular dependencies of the external radiation field, which is assumed to be isotropic in the rest frame of the supermassive black hole.} 
Such an additional photon field has also been shown to be necessary in the leptonic SED modeling of other ISP blazars~\citep{Boettcher:2013wxa}. 
Furthermore, \edit1{inclusion of external photon fields has been shown to significantly enhance the efficiency of HE neutrino production} \citep{Atoyan:2001ey,Murase:2014foa,Dermer:2014vaa}.}

\begin{figure}[th]
\centering 
\includegraphics[scale=0.45,trim=10 20 0 0]{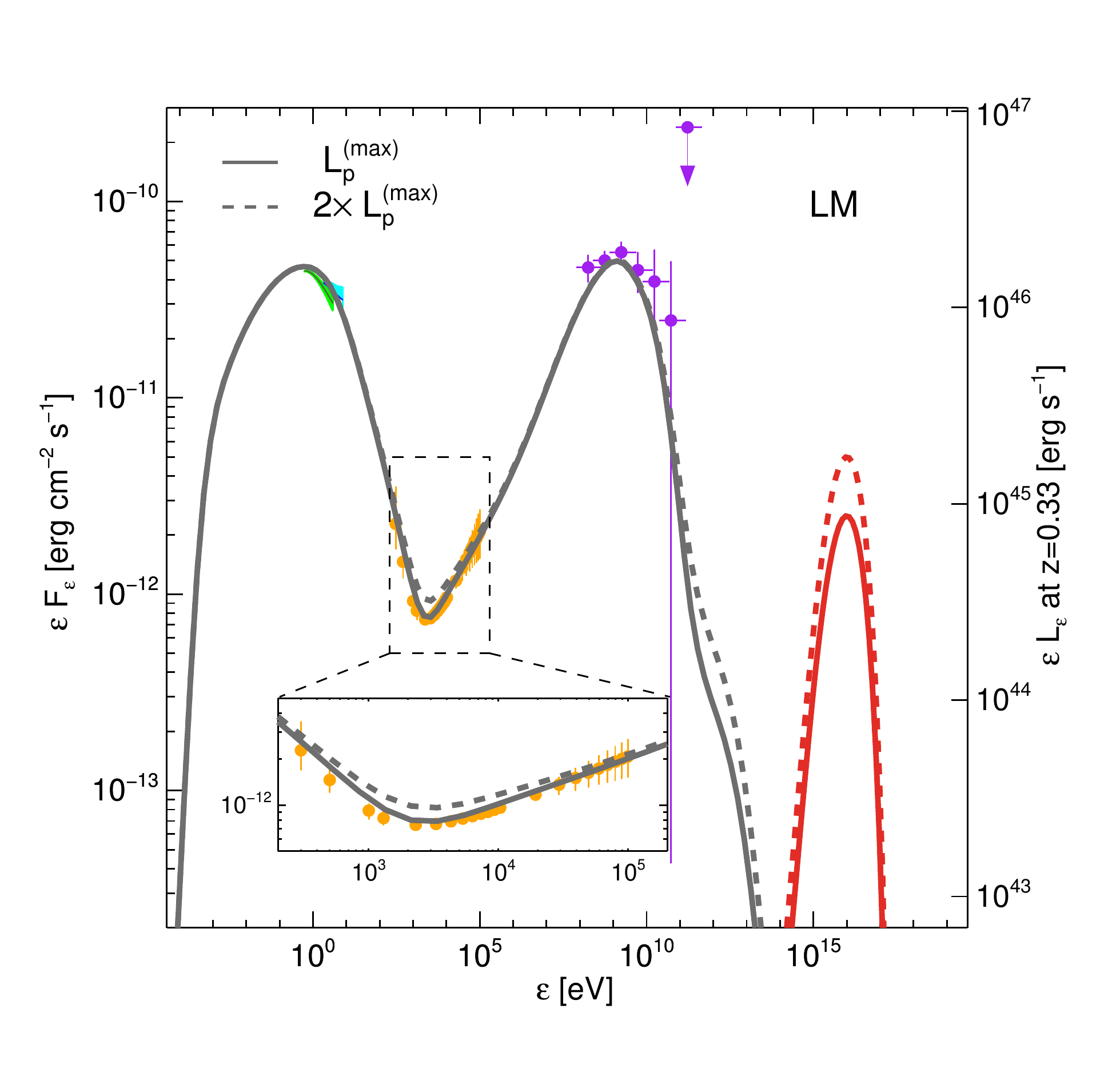}
\caption{Leptonic Model (LMBB2b) for the \txs\ flare (Ep.~1). Two SED cases (gray lines) are plotted against the observations (colored points, showing allowed ranges at 90\% confidence), one with hadronic component set to the maximum allowed proton luminosity $L_p^{(\max)}\approx 2\times 10^{50}$\,\ergsec\ (solid gray), and the other set to twice this maximal value (dashed gray line). Corresponding all-flavor neutrino fluxes for the maximal (solid red) and ``twice maximal'' (dashed line) cases are also shown.  Photon attenuation at $\egam\simgt3\times10^{11}$\,eV due to interactions with the extragalactic background light is not included here.
} 
\label{fig:leptonic}
\end{figure}

The respective photon spectrum and the maximum predicted neutrino flux for this parameter set (LMBB2b model) are presented in Fig.~\ref{fig:leptonic} (solid curves) and the parameter values are summarized in Tables~\ref{tab:tab1} and \ref{tab:tab1b}. We find that the \xray\ flux in the \nustar\ energy band is dominated by the SSC emission of the accelerated electrons, whereas the $\gamma$-ray emission is explained by the IC scattering of the fiducial external photon field by the same electron population. The steepening of the $\gamma$-ray spectrum at $\gtrsim 10$~GeV is due to the Klein-Nishina cross section. 
\edit1{Intriguingly, because of the steep \swift\ XRT spectrum and the low synchrotron peak-frequency revealed by our \xshooter\ data, the HE peak of the SED cannot be explained by the SSC emission alone.}
In addition, any attempt to describe the emission from a more compact region ($R^\prime \ll 10^{17}$~cm) fails because of the emergence of the SSC component which has a different photon index than the observed one in the \nustar\ band. 
This also demonstrates the importance of the detailed \xray\ data provided by this work. 

\begin{table}[bt]
\caption{Parameter values common to all leptonic models (LMs) for \txs}
\label{tab:tab1}
\begin{center}
\begin{tabular}{lc}
\hline 
$B^\prime$ [G] & 0.4  \\ 
$R^\prime$ [in cm] & $10^{17}$  \\
$\delta$ & 24.2 \\
$L^\prime_{\rm e}$ [in erg s$^{-1}$] & $2.2\times 10^{42}$ \\
$s_{e,1}$ & 1.9 \\
$s_{e,2}$ & 3.6 \\
$\gamma^\prime_{e, \rm min}$ & 1 \\
$\gamma^\prime_{e, b}$ & $5\times 10^3$  \\
$\gamma^\prime_{e, \rm max}$ & $8\times 10^4$ \\ 
\hline
\end{tabular}
\end{center}
\tablecomments{The isotropic-equivalent electron luminosity is $L_e=\delta^4L_e^\prime$. Parameter definitions are provided in Table~\ref{tab:parameters}.}
\end{table}

\begin{table*}[bt]
\caption{Model-specific parameter values for leptonic models (LMs) for \txs\ discussed in the text}
\label{tab:tab1b}
\begin{center}
\hspace*{-0.75in}%
\scalebox{0.75}{%
\begin{tabular}{l ccc ccc cc cc}
\hline
& LMBB1a & LMBB1b & LMBB1c & LMBB2a & LMBB2b & LMBB2c & LMPL1a & LMPL1b & LMPL2a & LMPL2b \\ 
\hline 
 $L_{\rm p}^{\prime (\rm max)}$ [$10^{44}$ erg s$^{-1}$] & 0.54 & 0.27 & 0.34 & 1 & 5.4 & 10 & 0.54 & 0.54 & 10 & 10\\
 $s_{p}$ & 2 &  2.5 & 3  & 2 & 2 & 2 & 2 & 2 & 2 & 2 \\
 $\gamma^\prime_{p, \rm min}$ & 1 & $3\times10^6$ & $3\times10^6$ & 1 & 1 & 1 & 1 & 1 & 1 & 1\\
 $\gamma^\prime_{p, \rm max}$ [$10^8$] & 30 & 30  &  30 &  1.6 & 0.16 & 0.016 & 30 & 30 & 0.016 & 0.016 \\
 $u_{\rm ext}^{\prime}$ [erg cm$^{-3}$] &  \multicolumn{6}{c}{0.033} & 0.033 & 0.067 & 0.04 & 0.08\\
 $T^\prime$ [K] &  \multicolumn{6}{c}{$3\times 10^5$} &  \multicolumn{4}{c}{n/a}\\
 $\alpha$ &  \multicolumn{6}{c}{n/a} & 3 & 2 & 3 & 2\\
 $\varepsilon_{\min}^\prime$ [keV] & \multicolumn{6}{c}{n/a} & \multicolumn{4}{c}{0.05}\\
 $\varepsilon_{\max}^\prime$ [keV] &   \multicolumn{6}{c}{n/a} &  \multicolumn{4}{c}{5} \\
 \hline 
\end{tabular}
}
\end{center}
\tablecomments{See Table~\ref{tab:parameters} for parameter definitions, and Table~\ref{tab:tab1} for parameter values common to all LMs. In LMBB models, the external photon field is blackbody-like with comoving temperature $T^\prime$, while in LMPL models, it is a power-law between comoving energies $\varepsilon_{\min}^\prime$ and $\varepsilon_{\max}^\prime$, with photon index $\alpha$. In all cases, $u'_{\rm ext}$ is the comoving energy density of the external photon field. Note that the isotropic-equivalent cosmic-ray proton luminosity is $L_p=\delta^4L_p^\prime$.}
\end{table*}

As noted in the previous section, HE photons produced via photohadronic interactions are attenuated in the source and induce an EM cascade whose emission should emerge in the \swift\ XRT and \nustar\ bands. 
As a result, the neutrino and proton luminosities are strongly constrained by the \xray\ data.
The photon spectrum obtained with $L_p=2\times L_p^{(\rm max)}$ already violates the observed \xray\ data points.  
In Fig.~\ref{fig:leptonic}, the upper limit on the all-flavor neutrino flux at the neutrino peak energy is $\eFenu^{(\rm max)}\sim(2-3)\times{10}^{-12}~{\rm erg}~{\rm cm}^{-2}~{\rm s}^{-1}$. 

In what follows, we show that our neutrino flux limits are fairly insensitive to the exact parameter values that may affect the photomeson production optical depth. 

\begin{figure}[tb]
 \centering 
 \includegraphics[width=0.5\textwidth]{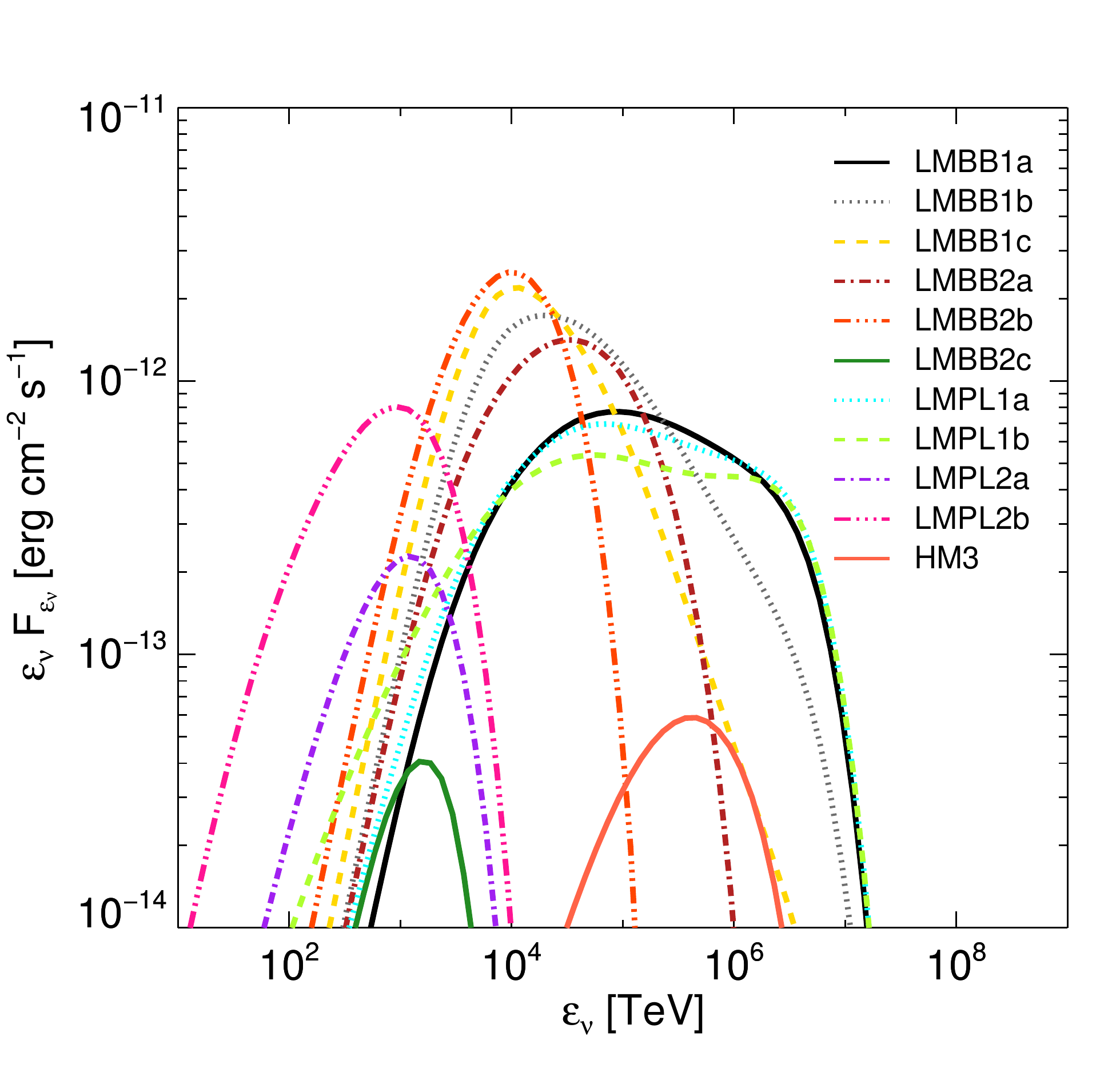}
 \caption{Upper limits on the all-flavor neutrino ($\nu+\bar{\nu}$) fluxes predicted for our modeling of the SED in the leptonic (LMx) and hadronic (HMx) models.} 
\label{fig:neutrino_all_sm}
\end{figure}

{\it Proton maximum energy ---}
\edit1{Motivated by the hypothesis that blazars are UHECR accelerators, i.e., at energies above $3\times{10}^{18}$~eV~\citep{2012ApJ...749...63M}, we explore the effect of the proton maximum energy on the neutrino flux upper limits. We thus explore cases with $\gamma'_{p,\max}=1.6\times{10}^8, 1.6\times{10}^9$, and $3\times{10}^9$ -- see Table~\ref{tab:tab1b}. Our results on the neutrino fluxes are presented in Fig.~\ref{fig:neutrino_all_sm}.}

Neutrino spectra in the \edit1{LMBB1x} models are more extended in energy compared to the default case (LMBB2b). They peak around 10~PeV (100~PeV) for $\gamma^\prime_{p,\max}=1.6\times10^7$  ($1.6\times10^8$) for LMBB2b (LMBB2a), respectively.
The number density of target photons decreases fast with increasing energy, \edit1{while} the photomeson production efficiency increases with energy~\citep{Murase:2015ndr}. 
However, the upper limits imposed on the proton luminosity and the peak neutrino flux are comparable in the LMBB2a and LMBB2b models. 
This is because the peak neutrino flux is bounded by the \xray\ data points through EM cascades, even though the photomeson production optical depths are quite different. 
As such, even lower maximum proton energies, e.g.\ $\gamma^\prime_{p,\max}=1.6\times10^6$, should not lead to higher upper limits on the neutrino flux. 
The reason is that protons with $\gamma^\prime_{p,\max}\sim10^6$ will produce electron-positron pairs (via the Bethe-Heitler process) on the synchrotron photons from the peak of the spectrum. 
Meanwhile, the photomeson interactions of the same protons on the \xray\ photons ($\nu\sim 10^{18}$~Hz) are less efficient~\citep{DMPR12}. 
The proton luminosity cannot be arbitrarily large in this regime, because the synchrotron emission from the Bethe-Heitler pairs will overshoot the \xray\ data.  


{\it Proton spectral index --- }
The slope of the power-law proton distribution is hardly constrained from the SED fitting.
Here, we investigate its effects on the neutrino spectrum by considering two additional cases with $s_{p}=2.5$ and $s_p=3$. 
For particle distributions with soft spectra (i.e., $s_{p}>2$), the total energy in protons is determined by the low-energy cutoff ($\gamma'_{p,\min}$) of the distribution. 
These low-energy protons, however, are of no interest for HE neutrino production. 
In an attempt to minimize the energy budget, while retaining the HE neutrino fluxes for $s_{p}>2$, one has to assume $\gamma'_{p,\min}\gg 1$ -- see Table~\ref{tab:tab1b}. 
The large $\gamma'_{p,\min}$ can also be justified if the proton distribution has a broken power law and the lower-energy segment has $s_{p}<2$ below the break (i.e. $\gamma'_{p, \min}\rightarrow\gamma'_{p,\rm br}$). 
Our results on the neutrino flux are presented in Fig.~\ref{fig:neutrino_all_sm} and compared to those obtained for $s_{p}=2$. 
The neutrino spectra become more sharply peaked as the proton distribution becomes softer, while the constraints on the 0.1--10 PeV neutrino flux \edit1{approach those of} our fiducial model (LMBB2b).
 

{\it External radiation spectrum --- }
Importantly, our results on the neutrino flux upper limit are insensitive to details of the unknown photon spectrum of external radiation fields. 
\edit1{In addition to the external} black body spectrum, we also consider a power-law spectrum. 
Such a broadband spectrum \edit1{might be produced, for example, by} electrons are accelerated with a hard spectrum in the sheath region of a structured jet, with the associated synchrotron photons -- with  a low synchrotron peak -- serving as seeds for the EIC emission in the \gray\ range~\citep{Tavecchio:2014eia,Tavecchio:2014xha}. 
From Fig.~\ref{fig:neutrino_all_sm}, we see that the results for \edit1{LMPL1x} models with $\gamma^\prime_{p,\max}=3\times10^9$ do not differ much from those for \edit1{LMBB1x} models. 
This is because the relativistic protons at ultrahigh energies mainly interact with target photons around the synchrotron peak. 
On the other hand, \edit1{LMPL2x} models with $\gamma^\prime_{p,\max}=1.6\times10^7$ give more optimistic neutrino fluxes than LMBB2c, because the photomeson interaction rate is enhanced compared to the photopair production rate \citep{petromast15}. 
However, the neutrino flux upper limits are still saturated at $\eFenu\sim10^{-12}$\,\ergcms, similar to that for LMBB2b. 
We thus conclude that the neutrino flux upper limit is $\eFenu^{(\rm max)}\sim{\rm a~few}\times10^{-12}$\,\ergcms \  whether the unknown target spectrum of the external radiation is described by a broadband power law or a narrower black body. 
In this work, we use the black-body spectrum as a fiducial case, which is conservative in the sense that it introduces fewer free parameters. 





In summary, in the leptonic models (LMs; Fig.~\ref{fig:leptonic}), the $\gamma$-rays are explained by the EIC emission, while there is a small contribution of the SSC component to the hard X-ray band. We note that the SSC component alone cannot explain the \gray \ component of the SED, \edit1{mainly because of (i) the separation of the low- and high-energy humps of the SED, (ii) the steep \swift\ XRT spectrum with the low synchrotron peak inferred by the \xshooter\ data, and (iii) the flat broad \fermi-LAT spectrum.}
Accelerated protons, generating HE neutrinos by photohadronic processes, are also present in this scenario, but with an associated EM component that is subdominant in $\gamma$ rays. 
The maximal all-flavor neutrino flux over ${\rm 0.1\,PeV}<\enu<{\rm 10\,PeV}$ is $\eFenumax\approx 3.6\times{10}^{-12}$\,\ergcms, implying a Poisson probability to detect one event with \ice\ over the six-month duration of the \txs\ \gray\ flare of at most $\sim1$\% under our assumed conditions, which are subject to model and observational constraints but \edit1{otherwise} optimal for HE neutrino production. 
See \edit1{Table~\ref{tab:tab2} and Sec.~\ref{sec:implications}} below for estimates of the expectation number of HE muon neutrinos for different model cases. 
The maximum proton isotropic-equivalent luminosity consistent with the SED is $L^{(\max)}_p\approx2\times{10}^{50}$\,\ergsec.
Cases with proton luminosities exceeding $L_p^{(\max)}$ lead to higher neutrino fluxes, but they are bounded by the observed \xray\ data due to electromagnetic cascade effects -- as shown in the inset plot, the ``twice maximal'' case already violates these constraints.  
We study different parameters to investigate the parameter dependence, and considered both black-body--like and power-law spectra for the external target radiation field.  
As a result, we find that the LM can provide at most a few percent expectation of an associated HE neutrino detection by \ice. 


\subsection{Hadronic Models (HMs)}
\label{sec:hadronic}

In hadronic scenarios, while the low-energy peak in the blazar's SED is explained by synchrotron radiation from relativistic primary electrons, the HE peak is explained by EM cascades induced by pions and muons as decay products of the photomeson production~\citep{mannheim93,muecke03}, or synchrotron radiation from relativistic protons in the ultrahigh-energy range~\citep{aharonian00,muecke03}. 
We coin this scenario ``HM'', which stands for {\sl Hadronic Model}, \edit1{in reference to} the hadronic origin of the \grays. 
The synchrotron and IC emission of secondary pairs may have an important contribution to the bolometric radiation of the source. 
In contrast to the leptonic scenario (Sec.~\ref{sec:leptonic}), the parameters describing the proton distribution can be directly constrained from the \nustar\ and \fermi\ LAT data. 
For the \txs\ flare, in the hadronic scenario, the SED can be fully explained without invoking external radiation fields.

There are different combinations of parameters that can successfully explain the SED in the HM scenario~\citep{boettcherreimer13,cerruti15}. 
As a starting point, we search for combinations of $\delta$ and $B^\prime$ that lead to rough energy equipartition between the magnetic field and protons, since the primary electron energy density is negligible in this scenario. 
With analytical calculations we derive rough estimates of the parameter values for equipartition: $\delta_{\rm eq}\sim 5$, $B^\prime_{\rm eq}\sim80$~G, $R_{\rm eq}^\prime\sim 10^{16}$~cm, and $\varepsilon^\prime_{p,\rm max}\sim 10^9$~GeV~\citep{petro16}. 

\begin{table}[bt]
\centering 
\caption{Parameter values for hadronic models (HMs) for \txs\ discussed in the text and presented in Fig.~\ref{fig:hadronic}.}
\label{tab:tab3}
\begin{center}
\begin{tabular}{l ccc}
\hline 
       & HM1 & HM2 & HM3\\
\hline
$B^\prime$ [G] & \multicolumn{3}{c}{85} \\ 
$R^\prime$ [in $10^{16}$cm] & 2  & 3 & 4.5\\
$\delta$ & 5.2 & 10  &  15  \\
$L_{e}^{\prime}$ [in $10^{43}$ erg s$^{-1}$] 9.3 & 0.6 & 0.06 \\
$s_{e,1}$ &  \multicolumn{3}{c}{1.8}\\
$s_{e,2}$ & 4.2 & 3.6 & 3.6   \\
$\gamma^\prime_{e,\rm min}$ [in $10^2$]& 6.3 &  1 &  1  \\
$\gamma^\prime_{e,\rm br}$ [in $10^2$] & 7.9 &  6.3  & 5 \\
$\gamma^\prime_{e,\rm max}$ & \multicolumn{3}{c}{$10^4$} \\ 
$L_{p}^{\prime}$ [in $10^{46}$ erg s$^{-1}$] & 2.7 & 0.1 & 0.01\\
$s_{p}$ & \multicolumn{3}{c}{2.1}\\
$\gamma^\prime_{p,\rm min}$ & \multicolumn{3}{c}{1}\\
$\gamma^\prime_{p,\rm max}$  & \multicolumn{3}{c}{$2\times10^9$}   \\
\hline 
\end{tabular}
\end{center}
\tablecomments{Parameter definitions are provided in Table~\ref{tab:parameters}.}
\end{table}

\begin{figure}[th]
\centering 
\includegraphics[scale=0.45, trim=10 20 0 0]{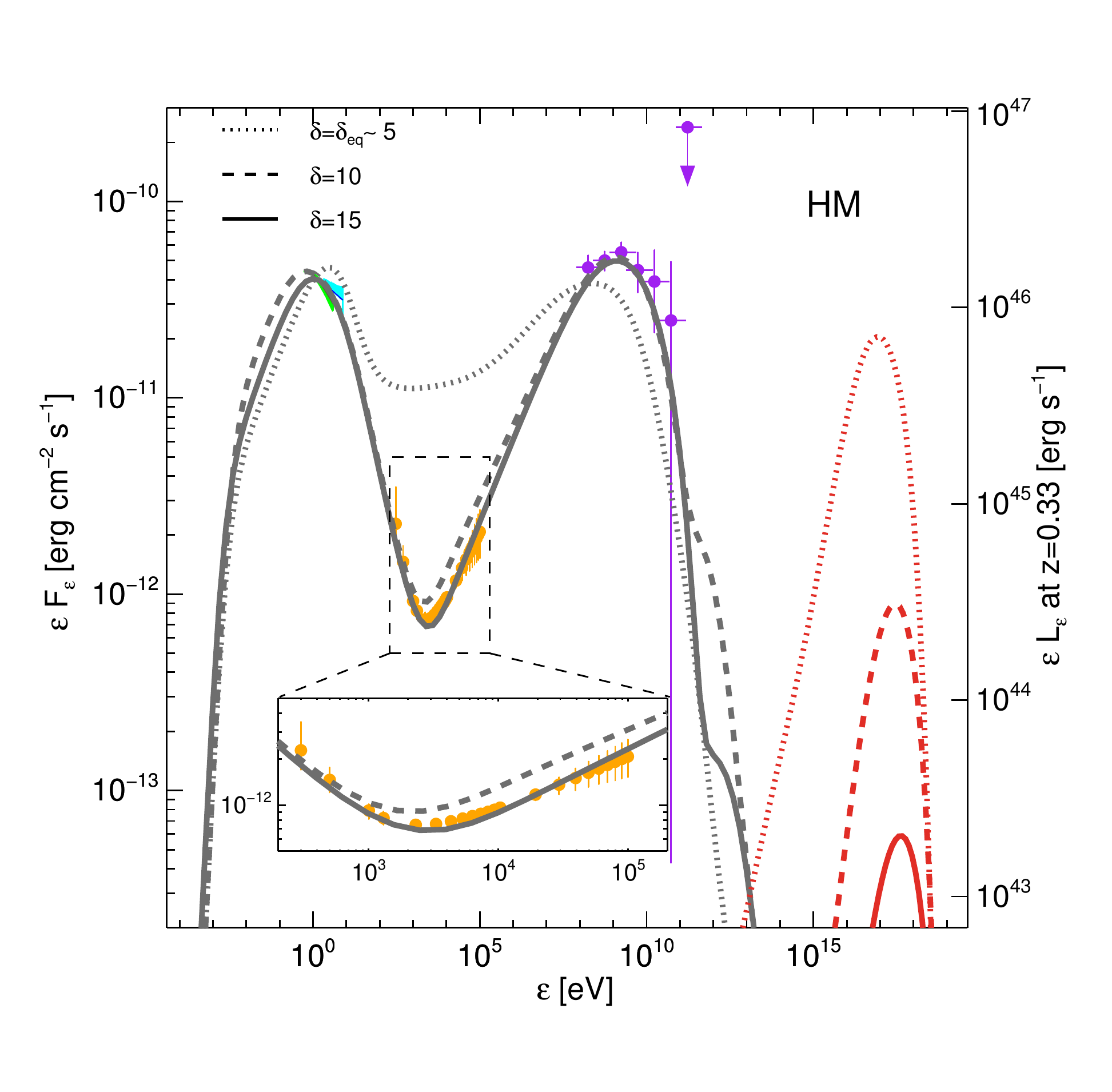}
\caption{Hadronic Model (HM3) for the SED of \txs\ flare (Ep.~1), as computed for different values of the Doppler factor (gray curves), together with resulting all-flavor neutrino fluxes (red curves) and electromagnetic observations (colored points, showing allowed ranges at 90\% confidence). Photon attenuation at $\egam\simgt3\times10^{11}$\,eV due to interactions with the extragalactic background light is not included here.}
\label{fig:hadronic}
\end{figure}

The parameter values obtained by numerically modeling the SED (see Fig.~\ref{fig:hadronic}) are summarized in Table~\ref{tab:tab3} and are similar to the estimates provided above. 
The jet power computed for this parameter set (HM1) is close to the minimum value expected in the hadronic scenarios. 
More specifically, the absolute power of a two-sided jet inferred for these parameters is $L_{j}\approx2\pi cR^{\prime 2}(\delta/2)^2 (u^\prime_{p}+u^\prime_{e}+u^\prime_{B})\sim4\times 10^{47}$~erg s$^{-1}$, with $u^\prime_{p}\approx2u^\prime_{B}\sim500$~erg cm$^{-3}$, where $u^\prime_{p}$, $u^\prime_{e}$, $u^\prime_{B}$ are comoving energy densities of relativistic protons, electrons, and magnetic fields, respectively. 
As demonstrated in Fig.~\ref{fig:hadronic}, the emission from the EM cascade forms a ``bridge'' between the low-energy and high-energy peaks of the SED for $\delta=\delta_{\rm eq}$ (gray dotted line). 
Despite minimizing the power of the jet, the adopted set of parameters for HM1 cannot explain the SED due to \edit1{the associated} significant EM cascade component. 


The EM cascade emission \edit1{can be} suppressed if the source becomes less opaque to the intra-source $\gamma\gamma$ absorption for HE photons.
This can be achieved for larger values of the Doppler factor since $\tau_{\gamma\gamma}\propto\delta^{-4}$ \edit1{(see also \citealt{murase16,Petropoulou:2017ymv}, for analytical expressions)}. 
The photon and neutrino spectra for $\delta=10$ and 15 are also shown in Fig.~\ref{fig:hadronic}, while the respective parameter sets (HM2 and HM3) are listed in Table~\ref{tab:tab3}. 
The SED is \edit1{compatible with} $\delta \ge3\delta_{\rm eq}$ (gray solid line). 
However, the photomeson production optical depth becomes lower as the two-photon annihilation optical depth decreases. 
In fact, the sub-PeV neutrino production efficiency is related to the opaqueness for $\gamma$ rays in the \fermi \ LAT band~\citep{murase16}.  
Furthermore, this model unavoidably leads to a higher jet power, i.e. $L_{\rm j}\sim6\times10^{48}$~erg s$^{-1}$, and $u^\prime_{\rm B}\gg u^\prime_{\rm p}$~\citep{petro17}. 
Moreover, as the Doppler factor increases, the peak of the neutrino energy spectrum is pushed to the EeV energy range \citep{DPM14}, while the neutrino flux in the 100~TeV--10~PeV range decreases due to the low efficiency of photomeson interactions.

HM3 demonstrates that the SED of the \txs\ flare can nicely be explained by the proton synchrotron model, but \edit1{with the consequence that} the HE neutrino production inside the source is very inefficient because of the \xray\ \edit1{constraints on} EM cascade emission. 
The acceleration of UHECRs with $\varepsilon_p\gtrsim3$~EeV is promising in this model, but cannot be reconciled with \edit1{an \icnu\ association}, since the predicted neutrino flux is too low in the 0.1--10 PeV energy range. 

In summary, we find that no reconciliation of the EM and neutrino observations is possible in hadronic models (HMs; Fig.~\ref{fig:hadronic}). 
The proton-induced cascade model that predicts $\eFegam\sim\eFenu$ unavoidably overshoots the observed \xray\ flux, giving $\eFegam\approx8\times{10}^{-12}$\,\ergcms, which is strongly excluded. 
Alternatively, the proton synchrotron model can explain the \txs\ \gray\ emission, but gives a maximal neutrino flux $\eFenumax\approx2\times{10}^{-15}$\,\ergcms, which implies a very low probability for \ice\ neutrino detection, $p_{\rm IC}< 10^{-5}$.
If the Doppler factor is sufficiently large, the proton-induced cascade emission is suppressed and can avoid overproduction of \xrays\ (see main and inset plots over 0.3--100\,keV), but at the price of a reduced neutrino flux; hence, only the low neutrino flux case (red solid curve) is viable. 
Such low neutrino-flux cases, leading to negligible HE muon neutrino detection probabilities, cannot accommodate \edit1{production of \icnu}. 


\subsection{Implications of \icnu}
\label{sec:implications}

Relativistic protons of energy $\varepsilon_{p}$ can interact with photons in the source and produce neutrinos with energy $\sim\varepsilon_{p}/20$.\footnote{Hadronuclear reactions such as proton-proton collisions in blazar jets are expected to be too inefficient compared to the photomeson production process, for typical values of the jets' plasma density \citep{Atoyan:2002gu,Murase:2014foa}. Also, even {\it ad hoc} high-density environments are similarly ($\sim$order of magnitude) constrained by the cascade bound.}
The targets for photohadronic interactions can be synchrotron and IC photons emitted by primary and secondary electrons as well as external photons to the source, if present. 
For a typical synchrotron spectrum around the SED peak, the rectilinear approximation around the $\Delta$ resonance is usually valid~\citep{Murase:2005hy,Murase:2014foa}, and the characteristic proton Lorentz factor interacting with photons of the frequency $\nu_{\rm syn}=\varepsilon_{\rm syn}/h$ is given by~\citep{Murase:2015ndr}
\begin{eqnarray}
{\gamma'}_{p,b}& \approx & 0.5 \ \delta \ \bar{\varepsilon}_\Delta{(\varepsilon_{\rm syn})}^{-1}{(1+z)}^{-1} \nonumber \\
& \approx & 1.3\times10^9\, \delta_1\, \nu_{\rm syn,14.5}^{-1}(1+z)^{-1},
\label{eq:gpth}
\end{eqnarray}
where $\bar{\varepsilon}_\Delta\sim0.3$~GeV is the resonance energy. 
The respective neutrino energy, in the observer's frame, is then given by ${\varepsilon}_{\nu,b}\approx 0.05\, {\gamma'}_{p,b}\, \delta\, m_pc^2/(1+z)$. 
This is also an estimate of the peak energy of the HE neutrino spectrum:
\eqb 
{\varepsilon}_{\nu,b}\approx0.05\, {\varepsilon}_p^b \simeq 600~{\rm PeV}~\delta_1^2\, \nu_{\rm syn,14.5}^{-1}{(1+z)}^{-2}.
\label{eq:Evth}
\eqe
Neutrinos with lower energies than $\varepsilon_{\nu,b}$ can still be produced by interactions of lower-energy protons with higher-energy photons. 
However, the neutrino flux at such lower energies is expected to be lower than the flux at $\varepsilon_{\nu,b}$ due to the decreasing number density of target photons, unless the proton distribution is a super-soft power law (i.e., $s_{p}\gg2$). 
Similar estimates can be derived in the presence of external radiation fields, as demonstrated in \citet{Murase:2014foa}.  
Note that the main target photons for the photomeson production process in the LMBB2 and LMPL2 models are photons with energies above the synchrotron peak, since even protons with $\gamma'_p \sim \gamma'_{p,\rm max}$ do not satisfy the photomeson production threshold for the peak synchrotron photons -- see Eq.~(\ref{eq:gpth}).

As described above, EM cascade emission induced by cosmic-ray protons \edit1{readily fills the dip} between the two peaks of the SED (keV to MeV energies). 
Thanks to the optical depth correspondence between photomeson production and two-photon annihilation, efficient production of HE neutrinos can only be achieved for conditions that lead to a stronger EM cascade emission inside the source. 
In particular, the proton-induced cascade model, where the \fermi\ LAT \gray\ data is \edit1{primarily} explained by the proton-induced cascade emission itself, is strongly ruled out. 
Although this model naturally predicts $\eFenu\sim\eFegam\sim{10}^{-10}$~\ergcms, which is consistent with the observation of \icnu\ \citep{Aartsen2018blazar1}, the EM cascade component typically has a broadband energy spectrum that extends \edit1{through} the \xray\ range. 
Thus, this model does not provide a proper description of the broadband SED. On the other hand, the proton synchrotron model can explain the blazar's SED, but the 0.1--10 PeV neutrino flux is predicted to be very low, $\eFenu\sim{10}^{-15}$~\ergcms.


\edit1{Fig.~\ref{fig:neutrino_all_sm} presents the upper limits on the all-flavor neutrino fluxes obtained in the leptonic and hadronic models for a wide set of parameters. For the maximum neutrino flux displayed in the figure,} 
$\eFenu\sim(2-3)\times{10}^{-12}~{\rm erg}~{\rm cm}^{-2}~{\rm s}^{-1}$, the \edit1{corresponding} muon neutrino fluence is estimated to be:
\begin{eqnarray}
\varepsilon_\nu^2\phi_{\nu_\mu}^{\rm (max)}&\simeq&1.6\times10^{-5}~{\rm erg}~{\rm cm}^{-2} \times \nonumber \\ & & {\left(\frac{\eFenu^{(\rm max)}}{3\times{10}^{-12}~{\rm erg}~{\rm cm}^{-2}~{\rm s}^{-1}}\right)}{\left(\frac{\Delta T}{0.5~{\rm yr}}\right)},\,\,\,\,\,\,\,\,\,\,\,\,\,
\end{eqnarray}
where $\Delta T\sim0.5$~yr is the flare duration and the flavor ratio is assumed to be $\nu_e:\nu_\mu:\nu_\tau\approx1:1:1$. 
Then, using the effective area for EHE real time alerts, ${\mathcal A}_{\rm eff}\sim{10}^6~{\rm cm}^2$ in the PeV range~\citep{IC_realtime}, the expected number of muon neutrinos is estimated to be:
\begin{eqnarray}
{\mathcal N}_\nu&\sim&(\varepsilon_\nu\phi_{\nu_\mu}^{\rm (max)}){\Delta \varepsilon_\nu}{\mathcal A}_{\rm eff}\simeq 0.02 \times \nonumber \\ & & {\left(\frac{\eFenu^{(\rm max)}}{3\times{10}^{-12}~{\rm erg}~{\rm cm}^{-2}~{\rm s}^{-1}}\right)}{\left(\frac{\Delta T}{0.5~{\rm yr}}\right)} {\left(\frac{{\mathcal A}_{\rm eff}}{10^6~{\rm cm}^{2}}\right)},\,\,\,\,\,\,\,\,\,\,\,\,\,
\end{eqnarray}
where the width of the neutrino spectrum is assumed to be $\Delta\varepsilon_\nu\sim\ln(10)$. We numerically confirm this analytical estimate for the different models presented in Fig.~\ref{fig:neutrino_all_sm}.    

Table~\ref{tab:tab2} summarizes our results on the upper limits obtained for the integrated all-flavor neutrino fluxes by modeling the flaring SED of \txs. 
Assuming a flare duration of $\Delta T=0.5$~yr, we also evaluate the expected number of muon neutrinos by using the effective area for EHE real time alerts~\citep{IC_realtime}, taking into account a correction due to the Earth attenuation toward the direction to \txs. 
The exact effective area may be slightly different for \icnu\ \citep{Aartsen2018blazar1}, but the results do not change within a factor of two. 
In the LMBB2b model or LMPL2b model, which give the most optimistic neutrino fluxes among our parameter sets, for example, the probability for \ice\ to observe one event that physically originates from the \txs~flare is evaluated to be $p_{\rm IC}\sim1$\% based on Poisson statistics.
This value is already achieved for the near-optimal case, but 
\edit1{we allow a factor of two uncertainty for several possible reasons.}
First, the duration of the HE neutrino flare may be a bit longer (although the confidence \edit1{of association would be} reduced for longer durations). 
Second, the EHE effective area for \icnu\ may be slightly different. 
Then, taking into account these variations as well as model uncertainties, we may regard the case for the proton luminosity of $2L_p^{(\rm max)}$ as the most conservative limit, which gives $p_{\rm IC}<2$\%.

\begin{table}[bt]
\caption{The upper limit on the all-flavor neutrino flux $F_{\nu}^{(\rm max)}$ for the different models that satisfactorily explain the flaring SED of \txs\ -- for details, see Sections~\ref{sec:leptonic} and \ref{sec:hadronic}.}
\label{tab:tab2}
\begin{center}
\begin{tabular}{lcc|c} 
\hline
  &\multicolumn{2}{c|}{$\int F_{\varepsilon_\nu}^{(\rm max)} d\varepsilon_\nu$ [erg cm$^{-2}$ s$^{-1}$]} & ${\mathcal N}_\nu$ \\
\hline 
    & 100 TeV - 1 PeV & 100 TeV - 10 PeV & $\leq10$ PeV \\
LMBB1a & $1.6 \times 10^{-14}$ & $4.5 \times 10^{-13}$ & $1\times{10}^{-3}$\\
LMBB1b & $5.2\times 10^{-14}$ & $1.7\times 10^{-12}$ & $4\times{10}^{-3}$\\
LMBB1c &  $9.1\times 10^{-14}$  & $2.7\times 10^{-12}$ & $6\times{10}^{-3}$\\
LMBB2a   & $4.5 \times 10^{-14}$ & $1.1\times10^{-12}$ & $3\times{10}^{-3}$\\ 
LMBB2b   &$1.8\times 10^{-13}$& $3.6\times10^{-12}$ & $8\times{10}^{-3}$\\ 
LMBB2c   &$2.5\times 10^{-14}$& $7.3\times10^{-14}$ & $2\times{10}^{-4}$\\
LMPL1a   & $3.1 \times 10^{-14}$ & $5.2\times10^{-13}$ & $1\times{10}^{-3}$\\
LMPL1b   & $9\times10^{-14}$ & $6.3\times10^{-13}$ & $1\times{10}^{-3}$\\
LMPL2a   &$2.5\times 10^{-13}$& $ 5.2\times10^{-13}$ & $5\times{10}^{-3}$\\ 
LMPL2b   &$1.2\times10^{-12}$&  $2\times10^{-12}$ & $1\times{10}^{-2}$\\ 
HM3   & $1.6\times 10^{-16} $ & $2\times 10^{-15}$ & $4\times{10}^{-6}$\\ 
\hline
\end{tabular}\\
\end{center}
\tablecomments{The reported values correspond to the neutrino fluxes integrated over 0.1--1 PeV and 0.1--10 PeV ranges. 
The last column shows the expected number of muon neutrinos below 10 PeV for a flare duration $\Delta T=0.5$~yr.}
\end{table} 


\section{Discussion}
\label{sec:discuss}
The contribution \edit1{of blazar jets} to the diffuse neutrino flux has been calculated based on both leptonic and hadronic scenarios for the observed \grays, \edit1{under the common assumption of a constant neutrino flux} \citep{Murase:2014foa,Tavecchio:2014xha,Tavecchio:2014eia,Padovani:2015mba,Rodrigues:2017fmu}.
\edit1{Most of the most optimistic scenarios for the diffuse neutrino emission from blazars (for a review, see \cite{Murase:2015ndr}) can now be constrained by \ice\ \citep[e.g.][]{Aartsen:2016prl, icrc2017}. 
In addition, all of the model-independent analyses} (stacking, multiplet, and auto-correlation analyses) have \edit1{disfavored} the blazar population as the dominant \edit1{($\sim100$\%)} origin of IceCube's neutrinos, implying that their contribution is less than \edit1{$\sim$3\% to 30\%} of the diffuse neutrino intensity in the 0.1--1\,PeV range \citep{Murase:2016gly,Aartsen:2014ivk,Aartsen:2017kru,Aartsen:2017mau}.
\edit1{Importantly, even if the blazar contribution to the diffuse neutrino flux is subdominant, the flaring blazar-associated neutrinos} are detectable with the current detector since the atmospheric backgrounds can be reduced by temporal and spatial coincidence. 
Although blazars \edit1{do not exhibit the extreme variability of} non-repeating transients like gamma-ray bursts and supernovae, they are \edit1{highly} variable on different timescales and across the EM spectrum \citep{Kataoka:2001zg,Abdo:2010rw,2012A&A...544A..80G,Sobolewska:2014mla}, and have \edit1{long been one of the most promising possibilities for HE neutrino production} \citep{Atoyan:2001ey,Dermer:2012rg,Dermer:2014vaa,Kadler:2016ygj,Petropoulou:2016ujj,turleyblazar,Halzen:2016uaj,Gao:2016uld}.

We presented the time-averaged data for Ep.~1 and Ep.~2 \edit1{and used the former for multimessenger modeling of the source SED.} 
Although \edit1{we have argued that this is a reasonable approach, it has} some limitations. 
\edit1{In particular, we cannot exclude the possibility that physical conditions of the source change drastically on a short time scales.} 
Although there is currently no evidence of such rapid variability, this could affect the results of our time-average SED modeling. 
\edit1{Nevertheless,} the EM cascade effects are inevitable and the resulting \xray\ component must appear. 
We \edit1{thus} expect that our conclusions \edit1{are robust,} since the upper limits on  neutrino fluence are basically \edit1{set} by the observed \xray\ fluence.  

We \edit1{modeled} the SED for Ep.~1 with a 30~d window. 
The \xray\ flux varies by a factor of 6 within 60~d, but the time-average SED for Ep.~2 does not differ \edit1{much} from that for Ep.~1. 
This indicates that the physical conditions are approximately similar during Ep.~1 and Ep.~2, which justifies the setup of our calculations. 
The flare duration that is relevant for estimates of the signal neutrino fluence is at least $\Delta T\sim60$~d and the \gray\ data suggest that a duration of $\Delta T\sim0.5-1~{\rm yr}$ is possible. 

\edit1{As noted above, it is the neutrino fluence (i.e., the product of the duration of neutrino emission and the neutrino flux) that matters in the calculation of the expected number of events.  
It is therefore likely that the expected number of events from a non-flaring blazar integrated over the IceCube lifetime is larger than the one expected from a flare. 
In our optimal case, if the SED shape and X-ray flux in the steady state remain similar to those in the flaring state, the expectation value of the number of muons that can be found in the eight-year point source analysis on upgoing muons is $\sim1$ event. Thus, time-averaged \xray\ fluxes had to be higher in the past to obtain $\sim10$ neutrino events. 
To properly address this question one would need to have a good description of the non-flaring SED of the source, especially in the \xray\ range. We plan to compare the flaring emission with the non-flaring emission of TXS 0506+056 in a dedicated future work (Petropoulou et al. 2018 in preparation).}

\edit1{Although our results on the neutrino flux upper limit in the LMs are insensitive to details of the external photon field, we briefly discuss possible origins of the external photons.} 
The typical photon energy of the external radiation field in the black hole rest frame is $\varepsilon_{\gamma,\rm ext}=3 k_B T^\prime/\Gamma=2.5\ T^\prime_{5.5}/\Gamma_{1.5}$~eV and its energy density is $u_{\rm ext}\approx u_{\rm ext}^\prime/\Gamma^2\simeq3.3\times10^{-5}\Gamma_{1.5}^{-2}$~erg cm$^{-3}$ \edit1{(see Table~\ref{tab:tab1b})}. 
The putative external photon field is compatible with scattered disk emission or soft emission from the sheath region of the blazar jet \citep{Dermer:2014vaa,Tavecchio:2014eia}. 
Any additional external component should not exceed the observed fluxes in the optical, UV, and \xray\ data. \edit1{Thus,} its luminosity should be $L_{\rm ext}\lesssim{10}^{45}$ to ${10}^{46}$\,\ergsec, \edit1{depending on the spectral shape}. 

In the scattered disk emission case, if there is a scattering region with the Thomson optical depth $\tau_T$ at radius $R$, the energy density of the scattered emission can be $u_{\rm ext}\lesssim(3-30)\times{10}^{-5}(\tau_T/0.1){(R/3\times10^{18}~\rm cm)}^{-2}$~erg cm$^{-3}$. 
Alternatively, dissipation in the sheath region of the jet may lead to electron acceleration and associated synchrotron emission with a peak energy of $\varepsilon_{\gamma,\rm ext}\sim 20 \ {(\Gamma_s/2)}^2\Gamma_{1.5}^{-1}$~eV and luminosity of $L_{\rm ext}\sim7\times{10}^{45}{(R/3\times10^{18}~\rm cm)}^{2}\Gamma_{1.5}^{-2}{(\Gamma_s/2)}^4~{\rm erg}~{\rm s}^{-1}$, where $\Gamma_s$ is the Lorentz factor of the sheath region. 

\edit1{Third, external photons can be provided by the possible broadline region,} and the energy density of the broadline region can be written as: $u_{\rm BLR}\approx 0.26\ f_{\rm cov}$~erg cm$^{-3}$, where $f_{\rm cov}$ is the covering fraction~\citep{ghisellini_tavecchio08}. 
\edit1{However, there are two drawbacks.} 
The lack of broadline signatures in the optical spectrum of \txs\ and other BL Lac objects suggests that such line emissions \edit1{are weak}. 
\edit1{Also, such emissions will only important when the blob is located in the broadline region.} 
If we follow the treatment in \citet{Murase:2014foa}, the observed \gray\ luminosity indicates that the broadline region is located at $R\lesssim{10}^{16}$\,cm, so that the typical emission radius would be larger than the radius of the possible broadline region. 

\edit1{Even though blazars like \txs\ could make a significant contribution to the diffuse neutrino flux, it is premature to extrapolate our findings from the modeling of the \txs\ flare to other blazars. 
A more dedicated study is left for future work, after a representative set of flares have been modelled individually~\citep[see also][]{Padovani:2015mba}. 
So far, the only other flare that was modelled with the same numerical code is the thirteen-day (2010) flare of the high-frequency peaked blazar Mrk~421 \citep{Petropoulou:2016ujj}. 
In that case, the available EM data could be explained with a higher neutrino-to-\gray\ luminosity ratio than we find here. 
If we were to take the neutrino-to-\gray\ luminosity ratio for the \txs\ flare as representative of flaring blazars, their contribution to the diffuse neutrino flux would be smaller than previous estimates.}   

Finally, as we emphasized, our theoretical interpretation is based on the single-zone model. 
In this assumption, the broadband EM and HE neutrino emissions are produced in the same localized region in the blazar jet. 
Efficient HE neutrino production requires large optical depths for photomeson production. This, in turn, implies that the emitting region is optically thick for HE \grays\ \citep{murase16}. 
This is especially the case for blazars that have soft photon spectra, in which the EM energy of the attenuated very HE photons will reappear at lower energies through an EM cascade. 
The tight constraints stemming from the large optical depth of a localized region could be alleviated, if the photon and neutrino emissions originate from different regions in the jet, as in multi-zone models \citep{Dermer:2012rg,Murase:2014foa,Dermer:2014vaa}. 
\edit1{Although detailed discussion must be deferred to future work (Murase \etal\ 2018 in prep.), the constraints from the EM cascade are unavoidable even for such multi-zone models, so that further {\it ad hoc} adjustments of the source parameters seem necessary.}


\section{Conclusions}
\label{sec:conclude}

\edit1{We have used the best available multiwavelength data to construct the broadband SED of \txs, over 10 orders of magnitude in photon energy from $\approx$1\,eV to $>$10~GeV, proximate to its likely ($\sim$3$\sigma$; \citealt{Aartsen2018blazar1}) emission of the high-energy neutrino \icnu. 
Working with this SED and the likely neutrino association, we have explored multimessenger models for \txs\ to evaluate whether this neutrino association is physically reasonable, and if so,} \edit2{under what} \edit1{conditions of the blazar jet and jet environment.}

\edit1{We find that} a leptonic scenario with a radiatively-subdominant hadronic component provides the only physically-consistent single-zone picture for this source's multimessenger (EM and neutrino) emissions.
If \icnu\ is associated with this flaring blazar, then physical conditions were close to optimal for neutrino production during its flare. 
\edit1{We find a maximal all-flavor neutrino flux over ${\rm 0.1\,PeV}<\enu<{\rm 10\,PeV}$ of $\eFenumax\approx (2-4)\times{10}^{-12}$\,\ergcms. 
The inferred ratio of proton to electron luminosities is large, $L_p/L_e \sim 250$ to 500, with smaller values prohibited because increased optical depth to $\gamma\gamma\rightarrow e^+e^-$ would suppress the observed \grays.}
Under these conditions, we find a probability of $p_{\rm IC}\approx 1\%$ to 2\% for \ice\ to detect an HE \edit1{muon} neutrino in real time at some point during the blazar's six-month flare. 

\edit1{Since the blazar's $>$GeV emissions are dominated by leptonic processes, and since EM cascades efficiently redistribute hadronic EM emissions across the spectrum, we find that the SED exhibits its greatest sensitivity to hadronic acceleration processes across its 0.1--100\,keV ``dip.'' 
Flux variations over this energy range are more likely to reflect the source's high-energy neutrino emissions than its GeV--TeV flux state. 
We thus find that, going forward, regular \xray\ monitoring of \txs\ and related blazars, in conjunction with continued monitoring by high-energy neutrino observatories, will provide a critical test of single-zone blazar models. Moreover, careful selection of temporal acceptance windows via \xray\ observations, as in \citet{turleyblazar}, will likely yield the most sensitive search for further multimessenger sources.}

\edit1{Finally, we find that under the observed flaring conditions, assuming the \icnu\ association holds,} \txs\ was not a significant UHECR accelerator. 
This is because a proton spectrum extending to $\epro\simgt3$\,EeV would yield a neutrino spectrum peaking above 100~PeV \citep{Murase:2014foa}; with the neutrino peak flux bounded by \xray\ observations via cascade effects, this would strongly suppress the 0.1--10 PeV neutrino flux (Fig.~\ref{fig:neutrino_all_sm}). 

It is possible that multi-zone models, which more readily decouple blazar EM and neutrino emissions, may ultimately be required to explain multimessenger observations of \txs\ and other blazars. 
\edit2{Independent of whether this particular source association holds,} our results demonstrate that detection of even one or two coincident neutrinos can grant us deep insight into a source, and should energize \edit2{future} \edit1{searches for further multimessenger sources.} 


\acknowledgements
\edit1{The authors thank the referee for valuable comments that helped to improve the manuscript.}
The authors thank the IceCube Collaboration for publicly distributing HE neutrinos in real-time, the \swift\ and \nustar\ teams for their rapid responses to our target of opportunity requests, and the \fermi\ collaboration for their publicly available data and analysis software. The authors acknowledge helpful discussions with L. Hagen, J. Charlton, and M. Eracleous. 
A.K., D.B.F., J.J.D., and C.F.T. acknowledge support from the National Science Foundation under grant PHY-1412633; A.K. acknowledges support from the National Aeronautics and Space Administration \swift\ Guest Investigator Program under grant NNX17AI95G. 
The work of K.M. is supported by NSF Grant No.\ PHY-1620777 and the Alfred P. Sloan Foundation. A.K. and K.M. gratefully acknowledges support from the Institute for Gravitation and the Cosmos at the Pennsylvania State University. 
J.K. and A.T. acknowledge support from NASA contract NAS5-00136.
P.A.E. and J.P.O. acknowledge support from the UK Space Agency. M.P. acknowledges support from the L.~Jr. Spitzer Postdoctoral Fellowship.
S.C. thanks the Centre National d'Etudes Spatiales (CNES) for support and funding.
Based on observations collected at the European Organization for Astronomical Research in the Southern Hemisphere under ESO program 099.D-0640(A).

Correspondence regarding this work should be sent to Kohta Murase (\emaillink{murase@psu.edu}) or Maria Petropoulou (\emaillink{m.petropoulou@astro.princeton.edu}) for theoretical interpretations, and Azadeh Keivani (\emaillink{keivani@psu.edu}) or Derek Fox (\emaillink{dfox@psu.edu}) for observations.

\edit1{\software{XSPEC~\citep{xspec}, {\scshape HEAsoft} (v6.22.1), \xshooter\ pipeline (v.2.9.3; \citealt{Goldoni2006, Modigliani2010}), \texttt{Reflex}~\citep{Freudling2013}, \texttt{Molecfit}~\citep{Smette2015,Kausch2015}, \fermi\ \texttt{Science Tools} (\url{https://fermi.gsfc.nasa.gov/ssc/data/analysis/software/}), {\sc SOPHIA}~\citep{SOPHIA2000}
}
}

\hyphenation{Post-Script Sprin-ger}





\end{document}